\documentclass[twocolumn,amssymb,amsmath,
nofootinbib,tightenlines,showpacs,floatfix,
superscriptaddress,aps,prb]{revtex4-1}
\usepackage{amsfonts}
\usepackage{txfonts} %otvr needed for not identical
\usepackage{amsbsy}
\usepackage{bm}%
\usepackage[colorlinks=true,linkcolor=blue,filecolor=blue,menucolor=yellow,urlcolor=blue,citecolor=blue,anchorcolor=blue]{hyperref}
\usepackage{graphicx}% Include figure files
\usepackage{times} 
\usepackage{dcolumn}% Align table columns on decimal point
\usepackage[style=base]{caption}
\usepackage[style=base,font=small]{caption} % EMfig Font change here
\usepackage[justification=justified]{subcaption}

\begin{document}

\title{Quasiparticle conductance in Spin Valve Josephson Structures} %otvc

\author{Evan Moen}
\email{moenx359@umn.edu}
\altaffiliation{Present address: Johns Hopkins University, 
Applied Physics Lab, Laurel, Maryland 20723} %otvnov
\author{Oriol T. Valls}
\email{otvalls@umn.edu}
\altaffiliation{Also at Minnesota Supercomputer Institute, University of Minnesota,
Minneapolis, Minnesota 55455}
\affiliation{School of Physics and Astronomy, University of Minnesota, 
Minneapolis, Minnesota 55455}

\date{\today}

\begin{abstract}

We study the quasiparticle current in clean ferromagnetic
Josephson structures of the form $S_1/F_1/N/F_2/S_2$,
where $S$, $F$, and $N$ denote superconducting,
ferromagnetic or normal layers respectively. %otv3
Our
focus is on the structure of the  conductance $G$  as
a function of bias $V$, emphasizing the
subgap region. We use
a fully self consistent numerical
method, coupled to a transfer matrix procedure to extract $G(V)$. 
We choose material parameters appropriate to experimentally
realized Co Cu Nb structures.  %otv3
We find a resonance peak structure as a function of
the intermediate layer thickness and of the misalignement angle 
$\phi$ between
$F$ layers. To understand this resonance structure, we develop
an approximate analytic method. For experimentally
relevant thicknesses, the conductance has multiple subgap
peaks which oscillate in position between low and critical
bias positions. These oscillations occur in both $\phi$ and
the layer thicknesses. We compare
our results with those obtained for
the spin valve structures $(F_1/N/F_2/S_2)$
and discuss the implications
of our results for the fabrication of spin %otv3 valve
Josephson
devices.

\end{abstract}

%\pacs{74.45.+c,74.78.Fk,75.75.-c}  %wv

\maketitle

\section{Introduction}
\label{JJintro}

%otv new paragraph below

In the continuing search for improved and more efficient
computing and memory devices involving spin manipulation\cite{tsyzu,esch}, many new
%EMpaper2 - moved citation to end of clause
possibilities
have opened up involving ferromanetic ($F$) superconducting ($S$) hybrid
structures. 
%
%The ability to manipulate spins while
%taking advantage of the dissipationless current is the basic advantage
%one is trying to exploit.
%EMpaper2 - dissipationless? Even considering the F layer components? low energy consumption is the point to make I think.
%
These devices allow for the manipulation of spin states while taking advantage of
the low power consumption in superconducting devices.
 The fundamental
phenomena are proximity effects arising from Andreev\cite{Andreev} reflection
together with, in hybrid samples % otv2 materials 
with a nontrivial magnetic %EMpaper3 materials
structure, the conversion of singlet Cooper pairs into odd-parity\cite{berezinskii,Halterman2002}
triplets, whose existence has been experimentally\cite{Keizer2006}
demonstrated. Triplet formation drastically
changes the proximity effect since the $m=\pm1$ 
pairs are transmitted through the $F$ electrodes
over a much larger length scale. Charge and spin transport are coupled
and can be reciprocally manipulated via spin transfer torques (STT) or 
applied fields. In structures involving two $F$ layers,
such as $S_1/F_1/N/F_2/S_2$ hybrids, the misalignement angle
$\phi$ between $F$ layer magnetizations can be changed experimentally\cite{alejandro}.
%EMpaper2 - split sentence
Thus the triplet formation, and hence the proximity effect, depends
drastically  on $\phi$ (such triplets are precluded when
$\phi=0$ or $\phi=\pi$) and the equilibrium and transport
phenomena involved can drastically depend on such manipulations.
Hence,  hybrid  structures involving either
spin valve $S_1/F_1/N/F_2$ or Josephson % otv3 like 
$S_1/F_1/N/F_2/S_2$
configurations continue to be the focus of
intensive work\cite{Satchell2018,birge2018,birge2012,birge,%otv2 new1?,others?,
Bergeret2017}. 
%EMpaper2 - moved citation to end of clause

There is growing interest in using Josephson junctions in digital electronics, 
%otv moved and rewritten
such as the Rapid Single Flux Quantum (RSFQ)\cite{RSFQ,RSFQ2,Manheimer} %otv2  
device, where information is stored and transmitted rapidly via the flux quanta. 
These devices could be
made to be more efficient if one were to trade its transistor components for magnetic memory elements %otv3
such as the spin valve ($F/N/F$)~\cite{Holmes2013}. %otv Because of this, 
%otv there has been a lot
%otv of interest in studying ferromagnetic Josephson structures such as the $S/F/S$ and $S/F/N/F/S$ structures.
Most of the focus of recent
work on Josephson ferromagnetic
junction  devices has been on the current-phase %otv2
 relationship\cite{Ryazanov2001,Kontos2002,birge2018,Maekawa,hvw15}. %otv3
%placeholderX}. 
% of the Josephson junction.
The relative phase of the superconductors, at zero current,
is either $0$ or $\pi$. For the $S/F/N$ Josephson structures, the equilibrium state can be changed between the $0$ and $\pi$ state by varying 
the thickness of the ferromagnetic layer~\cite{hv2004}  
or the relative magnetization angle~\cite{hvw15}. 
In addition, the critical current is also oscillatory with the layer thickness and exchange field strength~\cite{Buzdin1982,Buzdin1991}. 
However, there are other important aspects of the Josephson structure that are independent of the phase. 
In this paper, and following
up on previous work on the $F/N/F/S$ valve %EMnov added the valve description
configuration\cite{wvhg,Moen2018,condpaper,spinsplitpaper}, we focus on the quasiparticle
current in the Josephson structures.

%otv many edits/cuts below
%Josephson junctions are electronic devices with high magnetic sensitivity due to
% the Josephson effect. 
%otv3 sentence moved
The  most prominent phenomenon in Josephson structures is
of course the Josephson current, the existence of supercurrents in the presence
of a non-superconducting junction. %EMpaper2 - finishing the thought
In ordinary Josephson junctions
there are two Josephson effects: the DC effect and the AC effect~\cite{Josephson}.
In the DC effect, an applied DC current runs through the Josephson junction at zero bias, up to a critical value, via the tunneling of the Cooper pairs. 
The AC effect describes the AC current driven by an applied bias with frequencies in the GHz range for an applied bias of order $10~\mu$eV~\cite{Kittel}. 
%otv below
In general,
the Josephson current is not the only current that runs through a Josephson junction, as there is also the contribution of
normal electron transport. In the two-fluid model, %otv3
%awakward definition deleted where the Cooper pair
%electrons are considered to have completely separate transport from the normal electron transport, 
we can express the net current in a Josephson structure
using the Resistively and Capacitively Shunted Josephson (RCSJ) model~\cite{RCSJ}
as: %otv2
\begin{equation}
 \label{RCSJeq}
I=I_c \sin\left(\theta\right)+ I_{qp} + C \frac{dV}{dt} %otv2 =EMpaper3
\end{equation} 
This equation describes a resistive and capacitive circuit element running in parallel with a pure Josephson junction of tunneling Cooper pairs.
This is a non-linear equation and will result in a hysteresis~\cite{Orlando} in the current vs voltage (I-V) curves if the time scale of the RC element $\tau_{RC}=RC$ 
is greater than that of the Josephson junction $\tau_J={\Phi_o}/{2\pi}{I_c R}$ where $R$ is the normal resistance of the junction. 
The $I_{qp}$ term is known as the quasiparticle current and represents the contribution due to normal electron transport. %otv2 
It can be characterized via the conductance, $G(V)\equiv dI_{qp}/dV$. %otv3
One can measure the quasiparticle  current %otv2 $G V$ current 
by shunting the junction. 
This leaves a hysteretic I-V characteristic at very low biases 
where the ``capture'' current is small (and the minimum nonzero voltage is small) in the DC Josephson effect for decreasing current. 
In the case of a non-tunnel junction,
such as a clean or weak-link junction, there may exist unique subgap conductance features. A metallic weak-link is an $S/N/S$ structure in which the Josephson junction is separated
by a thin metal, sometimes the same material as the superconductor. 
For example, a point contact may be formed with one superconductor in contact with a superconducting substrate.
Another example is the microbridge, where a thin bridge is etched between two superconducting ``banks''~\cite{Orlando}. Although continuously connected,
the intermediate region in each case is considered a normal metal constriction. This is because the constriction is smaller than the coherence length ($\ell\ll \xi_0$) which
destroys superconductivity within the region. These constrictions are therefore studied in the dirty limit~\cite{Parks}. In the clean limit theory, 
the transport properties are not affected
by a constriction or by impurity scattering~\cite{Orlando}. 

%otv more edits below 
Here we wish to study the quasiparticle current in the clean limit 
for $S_1/F_1/N/F_2/S_2$ %otv3
structures using a self-consistent method which we present below.
For such structures, the standard DC Josephson current 
has a more complicated form than the simple $\sin(\theta)$ structure  
of Eq.~(\ref{RCSJeq}). Assuming a fixed phase difference $\theta$, the
structure of that term has been previously\cite{hvw15} studied. Here,
however, we are particularly interested in the subgap structure of the
quasiparticle current, as described by the %otv2 
conductance $G(V)$. %otv3 \equiv dI_{qp}/dV$. %otv2  term. % the quasiparticle current
%conductance $G V$ term, the quasiparticle current. %otvend
%EMpaper2 - no parenthesis on G(V) since it's the G * V term, but 6 one way...

In 1969, L. J. Barnes discovered multiple conductance peaks within the subgap bias region, using superconducting Nb  %otv3
point contacts~\cite{Barnes}. In these Josephson structures,
the subgap region is considered to be any bias below $2\Delta$ (or $\Delta_1+\Delta_2$ in the case of two different superconductors).  
Barnes found conductance peaks %EMnov using his name instead of a pronoun is generally more proper I think
%otv3 using niobium contacts repetition avoided
for values of the bias of approximately $eV=2\Delta/n$ where $n$ is an integer. 
This subgap structure (SGS) has since been verified in other experiments on metallic weak-link 
junctions~\cite{Hansen,Baturina,Giubileo}. 
In 1982, Blonder, Tinkham, and Klapwijk (BTK)
determined~\cite{btk} how Andreev reflections change the conductance features of an $N/S$ heterostructure where in the subgap
region ($eV<\Delta$ in this case) the conductance may be twice that of the normal conductance. For nonzero interfacial scattering, this leads to peaks in the conductance at
the critical bias ($eV=\Delta$). This peak represents the increase in energy needed for an electron in the normal metal to transport into the superconductor just
above the superconducting energy gap, where the density of states is the highest. For biases less than the gap potential, the right-moving electron will instead Andreev reflect
as a left-moving hole.
In 1983, Octavio, Blonder, Tinkham, and Klapwijk (OBTK) described the phenomenon known as multiple
Andreev reflection (MAR)~\cite{OBTK}. 
In a superconducting junction that is biased between the two superconductors, an electron leaving the left superconductor will gain in energy before impinging
on the right superconductor. If the energy is lower than the gap, it will reflect as a hole which then gains energy before impinging on the left superconductor. This process repeats
itself until the original electron has gained enough energy to escape the gap, making multiple reflections in the process. There is thus a peak in conductance when the number
of reflections $n$ times the bias applied $eV$ is equal to the energy gap $2\Delta$. OBTK went %otv3 
 on to describe 
the subgap structure in $S/N/S$ junctions via the MAR, although what they find are peaks in the {\it resistance} 
for non-zero temperatures and/or non-zero scattering at the $S/N$ interfaces.
One important distinction is the plane wave assumptions of the clean limit theory as
opposed to diffusive theory describing the weak-links. In Ref.~\onlinecite{OBTK}, plane waves were used %otv3
to describe the reflection coefficients at the $N/S$ interfaces, but no
interference of the reflected waves from each interface was included. %otv3
In general, the plane waves may interfere upon multiple reflections which would diminish the subgap structure. 
However, as with any junction, there are quantum resonance effects due to the finite thickness of the layers separating the superconductors. 
There are also\cite{lu2019} spin effects in the MAR spectrum. %otvnov2
In addition, OBTK assume %otvnov plural
 a non-self-consistent pair potential, and we have shown\cite{condpaper,spinsplitpaper} %in previous work %otv %EMnov - leave out 'in previous work', appears below
that a self-consistent pair potential is necessary
to accurately describe transport~\cite{wvhg,condpaper,spinsplitpaper,Moen2018}. 
Other theoretical work on MAR in weak-link metallic junctions~\cite{Kummel,Cuevas} has been in the dirty limit~\cite{Anderson,Gorkov,Usadel}. 
We study these reflections and the resulting interference and resonance phenomenon
in our ballistic, self-consistent theory for the ferromagnetic Josephson structure.

In %otv2the 
previous work\cite{condpaper,spinsplitpaper} we have studied the quasiparticle transport in superconducting spin valve structures ($F_1/N/F_2/S$). 
In these structures, the
singlet Cooper pair correlations are short-ranged and oscillatory within the ferromagnet~\cite{Buzdin1990, Halterman2002}. The 
presence of a second ferromagnet allows for the formation of induced same-spin triplet correlations of the Cooper pairs which are long ranged within the 
ferromagnet~\cite{bvermp,Eschrig2008,leksin,Bergeret2007,kalcheim,singh,ha2016}.
Due to this, and the oscillatory nature of the singlet pair, 
we found that the subgap features of the system are highly dependent on the magnetic misalignment angle $\phi$ 
and the thickness of the $F_2$ layer. In Ref. \onlinecite{condpaper} we found that the critical bias (CB), 
i.e. the bias value equal to the saturated pair potential $eV=\Delta$, 
was spatially and angularly dependent.
In Ref. \onlinecite{spinsplitpaper} we saw that the conductance 
features are spin-split between contributions from %otv3
incoming spin-up and spin-down electrons 
where, in the subgap region, one spin-band features a peak in conductance while the other spin-band has a minimum. This lead to a peak conductance that is oscillatory with $V$ %otv3
between zero and the critical bias. %otv3
It was also shown that these conductance features were highly dependent on the interfacial scattering. In fact,
nonzero scattering is paramount to the formation of conductance peaks. These dependencies also apply to the $S_1/F_1/N/F_2/S_2$ system, and we will study
the thickness and angular dependence in the results of this work.

In Sec.~\ref{JJmeth}, we review our methods, which are the same as those used in 
Refs. \onlinecite{condpaper} and \onlinecite{spinsplitpaper}, to study the $S_1/F_1/N/F_2/S_2$ spin valve Josephson structure. In that section, we also review
our analytic approximation of the system to determine the relationship of the electron-hole resonance in $N/F/S$ and $S/F/S$ multilayers with interfacial scattering 
due to a normal metal contact. In Sec.~\ref{JJresults} we present our results, starting with our approximate %otv3
analytic calculations on the simple $N/F/S$ and $S/F/S$ models,
before moving on to the fully self-consistent, numerical calculations of the $S_1/F_1/N/F_2/S_2$ heterostructure. In all our calculations, we determine the thickness
dependence of the $F$ (or $F_2$) layer in relation to the resonance effects determined in Sec.~\ref{JJanalyticSub}. In addition, we determine the angular dependence
for our numerical calculation. We find that the angular dependence is 
different from %otv3 unique???unique to 
that found in the $F_1/N/F_2/S$ systems previously studied.
We present our results for two sets of interfacial scattering parameters: clean interfaces and imperfect $N/F$ interfaces. We also consider 
%EMpaper2 - not all of them have the X/S barrier now
nonzero scattering due to a normal metal contact. Finally, we summarize our results in Sec.~\ref{JJconc}.

%%%%%%%%%%%%%%%%%%%%%%%
\section{Methods}
\label{JJmeth}
%otv3 caption rewritten because figure is very different from that in thethesis. 
%otv3 we probably should repalce the figure?
\begin{figure}[ht] 
\centering
\includegraphics[width=0.45\textwidth] {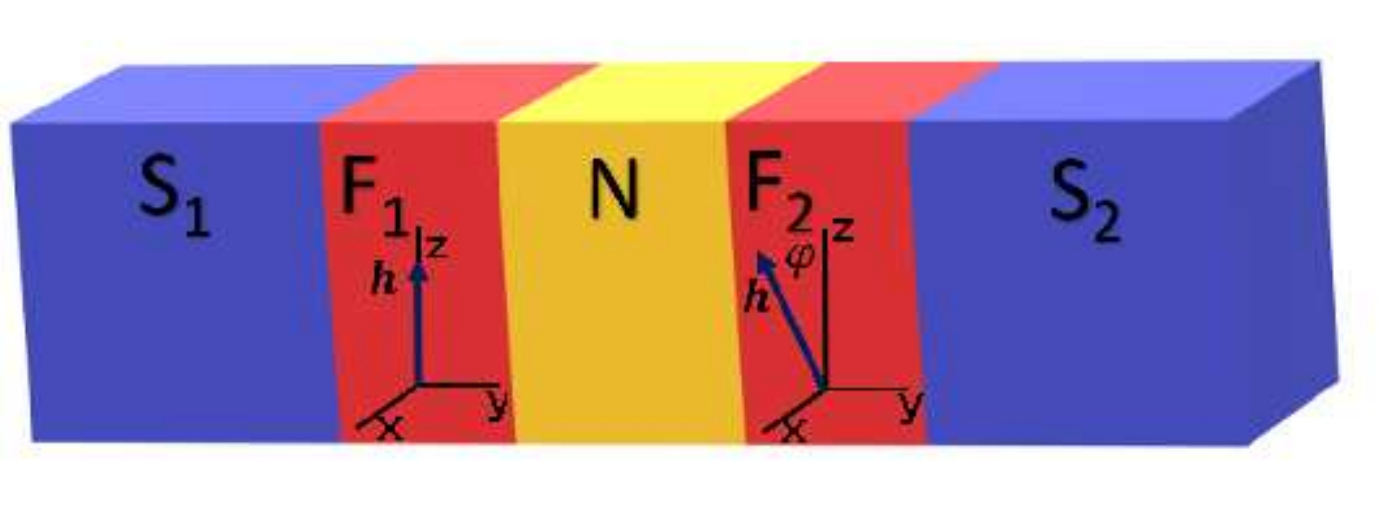}
\caption{Sketch of the $S_1/F_1/N/F_2/S_2$ heterostructure. 
%In the text, 
The $z$ axis is taken along the magnetization
of the outer magnetic layer $F_1$, 
%is along the $z$ axis 
while that in
$F_2$ it is rotated in the $x$-$z$ plane by an angle $\phi$. The $y$ axis is 
normal to the layers.
This sketch is not to scale. 
}
\label{JJfig0}
\end{figure}

\subsection{Self-consistent calculation of the pair potential} 
\label{pairpotSub}

%otv minor edits and referencing below
The methods in this section are very close to those used in 
the Refs.~\onlinecite{wvhg,hvw15,condpaper,spinsplitpaper} (and
references therein) 
 for the calculation of the
pair potential, and those used in Refs.~\onlinecite{wvhg, condpaper,spinsplitpaper} 
for calculating the conductance. 
The primary %otv3 moved
difference %in the calculation of the conductance %otv3 why conductance only?
is the inclusion of the second superconducting layer. 
The geometry we consider is depicted in Fig.~\ref{JJfig0}.
The layers are assumed to be infinite in the transverse direction ($x$-$z$ plane)
and the quasiparticle current is along the $y$ axis. %otv3 %EMnov - it sounded like the y-axis was infinite
The eigenvalue equation %otv
 for the $S_1/F_1/N/F_2/S_2$ multilayer, in
the quasi one-dimensional geometry considered here %otv %EMnov - 'one'
 is:
\begin{align}
&\begin{pmatrix}
{ H}_0 -h_z&-h_x&0&\Delta \\
-h_x&{ H}_0 +h_z&\Delta&0 \\
0&\Delta^\ast&-({H}_0 -h_z)&-h_x \\
\Delta^\ast&0&-h_x&-({H}_0+h_z) \\
\end{pmatrix}
\begin{pmatrix}
u_{n\uparrow}\\u_{n\downarrow}\\v_{n\uparrow}\\v_{n\downarrow}
\end{pmatrix} 
=\epsilon_n
\begin{pmatrix}
u_{n\uparrow}\\u_{n\downarrow}\\v_{n\uparrow}\\v_{n\downarrow}
\end{pmatrix}\label{bogo4}
\end{align} 
where $H_0=-(1/2m)(d^2/dy^2)+\epsilon_\perp -E_F(y)+U(y)$ is the usual single particle Hamiltonian with interfacial scattering $U(y)=\sum_{i} H_i\delta(y-y_{i})$
where $H_i$ is the barrier strength at the $i$th interface located at $y_i$.
Also, $\epsilon_\perp$ is the transverse kinetic
energy, %otv
$h$ is the exchange field within the ferromagnetic layers, and $\Delta$ is the pair potential within the superconducting layers. Each element in
the matrix equation is implicitly a function of the position ($y$) within the multilayer. 
The form of the Hamiltonian is the same as was given in previous
work\cite{wvhg,hvw15,condpaper,spinsplitpaper} and, because
of the presence of two superconductors we, must keep
in mind that\cite{hvw15} the pair potential
is not necessarily real: there may exist a phase difference %otv3
%otv Sec.~\ref{introBdG} except 
between the two $S$ layers, and we have therefore introduced the complex conjugate of 
the pair potential $\Delta^\ast$. With a single superconductor, there is only 
one phase associated with the s-wave symmetry and thus the
 pair potential can be taken to be real. %otv
Using our self-consistent method, as described
in the above references, we initialize the pair potential within each layer to a selected starting phase difference, $\Delta_1(y)=\Delta_0$
and $\Delta_2(y)=\Delta_0 e^{i \theta}$ where $\Delta_1$ is the value of 
$\Delta(y)$ for $y$ values within the $S_1$ layer, and similarly for $\Delta_2$.
We assume here that the two superconducting layers are made of the
same material. We %otv3  
then solve the eigenvalue Eq.~(\ref{bogo4}) and %otv3
then evaluate the self consistent equation,
\begin{equation}
 \Delta(y) = \frac{g(y)}{2}{\sum_n}^\prime\bigl[u_{n\uparrow}(y)v_{n\downarrow}^{\ast}(y)+
u_{n\downarrow}(y)v_{n\uparrow}^{\ast}(y)\bigr]\tanh\left(\frac{\epsilon_n}{2T}\right)
\label{JJselfconsistency}
\end{equation}
where $g(y)$ is zero {\it except} in the $S$ layers and the sum is over all eigenstates with energies less than $\omega_{D}$ from the Fermi level. 
We iteratively solve for $\Delta(y)$ 
by cycling through 
Eqs.~(\ref{bogo4}) and (\ref{JJselfconsistency}) %otv3.%otv as explained in  Sec.~\ref{introBdG}. 
The phase of the complex pair potential will also iterate using this method. For the equilibrium calculation (zero current), there are always
two local stabilities in the phase: $0$ and $\pi$. For an initial guess where the phase difference is not equal to $0$ or $\pi$, the
final self-consistent phase will always converge to the value %otv3
 which minimizes the free energy. This value %otv3  phase which minimizes the free energy 
is dependent on the thickness~\cite{hv2004} and relative magnetization angle~\cite{hvw15} 
of the ferromagnets. 
In non self consistent
calculations (such as our approximate
analytic calculations below), the $0$ and $\pi$ phases are degenerate, and thus we leave the phase to be zero. %otv should we move this?
In our numerical, self-consistent results presented, 
the overall phase corresponding to the plot displayed is that which minimizes the free energy, i.e. the equilibrium phase. 
The self-consistent method is  %otv3 made 
necessary to preserve 
the fundamental property of charge 
conservation\cite{bagwell,sols2,sols,baym}, 
as explained in Ref.~\onlinecite{condpaper}. %otvSec.~\ref{conservation}.

Using this method, it is possible to allow for two different superconductors ($\Delta_{0,1} \neq \Delta_{0,2}$) 
where the coherence length is different for each $S$ layer. 
We focus our attention here to the case in which the two superconductors are made of the same material. 
This does not mean that the pair potential $\Delta(y)$ will be symmetric or have the same
magnitude in each $S$ layer, as the proximity effect does not impact both superconductors equally as we vary the $F$ layer thicknesses. 
%otv33It should be noted that one could achieve 
The same 
asymmetry occurs if one varies the thickness of the individual $S$ layers. %otv3

\subsection{Quasi-particle Conductance}
\label{JJqpCondSub}

In the previous subsection, we reviewed the %otv
determination of the equilibrium properties of the $S/F/N/F/S$ system, 
in particular the spatial dependence of the pair potential. Now
we explain our methods for calculating the conductance. %otv which are 
% essentially the same as those presented in Refs.~\onlinecite{condpaper} and \onlinecite{spinsplitpaper}. 
Basically, 
we use here the same BTK~\cite{btk} method to calculate the conductance as %otv
 in previous work\cite{condpaper,spinsplitpaper}. Hence, we will
not repeat any of the details here. In the BTK method one %otv
calculates the conductance from
the reflection amplitudes of the spin-dependent ordinary ($b_\sigma$) and 
Andreev ($a_\sigma$) reflection amplitudes within the left-most layer.
In order
to simplify our calculations, 
we introduce a thin normal metal contact, denoted by '$X$', located to the left 
of the $S_1$ layer (see Fig.~\ref{JJfig0}). 
This makes the ferromagnetic Josephson structure effectively $X/S_1/F_1/N/F_2/S_2$. 
This contact layer is taken to be thin enough to not affect the calculation
of the pair potential through the proximity effect. For this reason, it is not necessary to include it in the equilibrium calculation above. We further explain
our reasoning for introducing this $X$ contact below.
 %otv Eq.~(\ref{conductance_cond}).
We determine the reflection amplitudes by writing the incoming wavefunctions using these amplitudes and applying the boundary conditions at the
end layers and the continuity conditions at the interfaces. If a spin-up incoming electron in the left-most layer is traveling in the normal metal contact $X$, 
the incoming wavefunction is:
\begin{equation}
\label{Xwaveup}
\Psi_{X,\uparrow}\equiv\begin{pmatrix}e^{ik^+_{N}y}+b_{\uparrow,\uparrow}e^{-ik^+_{N}y}
\\b_{\downarrow,\uparrow}e^{-ik^+_{N}y}
\\a_{\uparrow,\uparrow}e^{ik^-_{N}y}
\\a_{\downarrow\uparrow}e^{ik^-_{N}y}\end{pmatrix}\enspace\enspace
\end{equation}
and for a spin-down incoming electron:
\begin{equation}
\label{Xwavedown}
\Psi_{X,\uparrow}\equiv\begin{pmatrix}b_{\uparrow,\uparrow}e^{-ik^+_{N}y}
\\e^{ik^+_{N}y}+b_{\downarrow,\uparrow}e^{-ik^+_{N}y}
\\a_{\uparrow,\uparrow}e^{ik^-_{N}y}
\\a_{\downarrow\uparrow}e^{ik^-_{N}y}\end{pmatrix}\enspace\enspace
\end{equation}
where the second spin index of the reflection amplitudes denotes the spin of the incoming particle and 
$k^{\pm}_{N}=\left[E_F\pm{\epsilon}-{k_\perp^2}\right]^{1/2}$ is the normal metal wavenumber. 
%otv repetitive? The procedure follows that used in Refs.~\cite{condpaper}
%and \cite{spinsplitpaper}. 
The continuity condition of the wavefunctions at each interface can be represented by a matrix equation 
$M_{i,r} x_i = M_{i+1,\ell} x_{i+1}$ where $M_{i,r}$ and $M_{i,\ell}$ are the wavefunction coefficients of the $i$th layer evaluated 
at the right and left interface respectively, and $x$ is the vector of the
reflection/transmission amplitudes. At the $X$ layer, this equation becomes $M_X x_X  + c_\sigma = M_{S1,\ell} x_{S1}$ where $c_\sigma$ is the vector of the 
incoming spin $\sigma$ electron coefficients. The wavefunctions are described in Ref.~\onlinecite{condpaper} for the ferromagnetic ($F$) and superconducting ($S$) layers.
The addition of a second superconducting layer is straightforward as 
one uses the same self-consistent approach as for one
$S$ layer (see Ref.~\onlinecite{condpaper}). %EMpaper2 - parenthetical

The conductance is then calculated via the BTK method, in the low $T$ limit, %otv3 
 using (see
e.g. Eq.~(10) of Ref.~\onlinecite{spinsplitpaper}): %otv3 %Eq.~(\ref{conductance_cond}).
\begin{align}
\label{conductance_cond}
&G(\epsilon)=\sum_\sigma P_\sigma G_{\sigma }(\epsilon) %otv3 not G_0
\\\nonumber
&=\sum_{\sigma}P_{\sigma}\left(1+\frac{k^-_{\uparrow 1}}{k^+_{\sigma 1}}|a_{\uparrow,\sigma }|^2
+\frac{k^-_{\downarrow 1}}{k^+_{\sigma 1}}|a_{\downarrow,\sigma }|^2
-\frac{k^+_{\uparrow 1}}{k^+_{\sigma 1}}|b_{\uparrow,\sigma }|^2
-\frac{k^+_{\downarrow 1}}{k^+_{\sigma 1}}|b_{\downarrow,\sigma }|^2\right), %wv 
\end{align} 
where the index $1$ refers to the leftmost layer. $G$ is given 
in the customary  
natural units of conductance $(e^2/h)$.
The factors
$P_\sigma$ take into
account any possible different density of incoming spin up and spin down states.
The energy dependence of $G(\epsilon)$ arises from the applied bias voltage $V$.
We measure this bias in terms of the dimensionless quantity 
$E\equiv eV/\Delta_0$ where $\Delta_0$ is the value of the order parameter in
bulk $S$ material. 
In general one has for the wavevectors:
\begin{equation}
\label{wavevector_cond}
k^{\pm}_{\sigma 1}=\left[(1-\eta_{\sigma}{h}_1)\pm{\epsilon}-{k_\perp^2}\right]^{1/2},
%\end{align}
\end{equation}
where $\eta_\sigma \equiv 1(-1)$
for up (down) spins, and $k_\perp$ is the wavevector %
corresponding to energy $\epsilon_\perp$ and $h_1$ the internal
field of the leftmost layer. All
wavevectors are in units of $k_{FS}$ and all
energies in terms of $E_{FS}$.
The presence of the  normal metal contact $X$ %otv 
means that we can describe the left-most layer using incoming electrons
and holes, as opposed to the electron-like and hole-like quasiparticles of the superconductor 
(see e.g. Ref.~\onlinecite{btk}). Hence $k^{\pm}_{\sigma 1}$ 
are  actually
spin independent, %otv3  and equal to $k^{\pm}_{N}$. k_N undefined?
and $P_\sigma=1/2$. %otv3
Although a description using incoming electron/hole-like quasiparticle amplitudes has been used in studies on the phase relationship in Josephson structures~\cite{Furusaki}, 
this approach is not well suited for determining the transport properties within the subgap region as these amplitudes can not describe excitations with a subgap energy -- 
only energies above the gap. 
To probe the subgap energies means describing incoming Cooper pairs instead of the excitation amplitudes. 
With the $X$ layer we are able to describe the quasiparticle states 
for the subgap in terms of the incoming electron/hole excitation amplitudes. 
In addition, in the BTK method  (see Eq.~(\ref{conductance_cond})), %otv3
$G$ is described via the reflected electron
and Andreev reflected hole amplitudes. To describe the system using the electron-like and hole-like quasiparticle reflection coefficients would 
require an entirely new formalism. 
Adding a normal metal contact 
is also justified on the basis that experimental systems have contacts from which measurements are made. In addition, we can study the effects of the interfacial
scattering due to imperfect contact interfaces. The introduction of a scattering interface allows for multiple Andreev reflections in both single superconductor
heterostructures and the Josephson structures, the results of which are conductance peaks in the subgap region sometimes known as the subgap structure (SGS).

%otv In the next subsection,
%we go over our analytic approach to better understand the origins of the subgap %conductance peaks.

\subsection{Analytic Approximation}
\label{JJanalyticSub}

\begin{figure}[ht]
\includegraphics[angle=-90,width=0.45\textwidth] {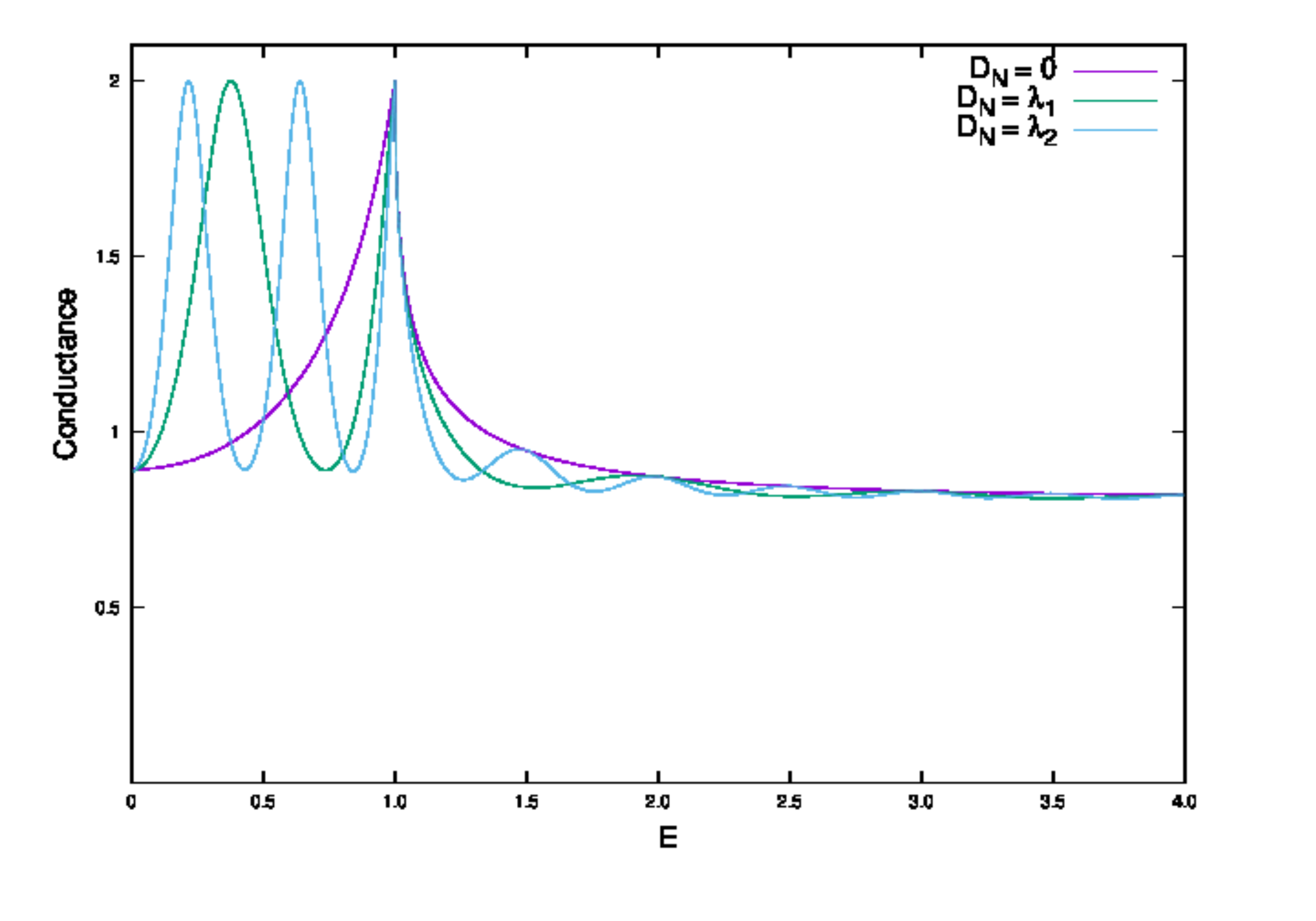} 
\vspace*{-0.2cm}
\caption{Conductance (G) vs. Bias (E) in the $N^\prime/N/S$ structure for varying $D_N$. $\lambda_n = \frac{\pi^2}{2}n\Xi_0$ 
are the resonance values at which a new peak forms at the critical bias, moving the previous peaks into the subgap region. 
We assume a single barrier at $N^\prime/N$ interface $H_B=0.5$. Analytic approximation. 
}
\label{NSCfig1_JJ}
\end{figure}

%otv some edit below
We use also in this paper an analytic approximation. Its purpose
 is to provide some physical intuition and a
quantitative description of the finer details in the full
numerical $S_1/F_1/N/F_2/S_2$ results. To do this we start with a simple $N/F/S$ model (see Figs.~\ref{NSCfig1_JJ} and \ref{NSCfig2_JJ}) and go on
 to a ferromagnetic
Josephson structure $S_1/F/S_2$ with normal metal contact $X$ (see Figs.~\ref{NSCfig3_JJ} and \ref{NSCfig4_JJ}). 
To make our calculation analytic, we need to make many approximations: 
we assume a one-dimensional (as opposed to quasi-one dimensional) 
system with infinite layer 
thicknesses at the left and right ends. Therefore, the only thickness dependencies come from the intermediate layers $F$ and $S_1$. 
We must assume also
a non-self consistent pair potential where $\Delta_1=\Delta_2=\Delta_0$ is 
a constant for both the single $S$ and Josephson structures. 
%otv repeated? In the case of the ferromagnetic Josephson structure, we include a normal metal contact $X$ to the $S_1$ layer
%with nonzero interfacial scattering at the $X/S_1$ contact interface, for reasons described above. 
The calculation of the conductance is then a simpler version of that in the numerical calculation.
The reflection amplitudes can be solved for %otv
using $x_N=\mathcal{M}^{-1}_N\mathcal{M}_{F,\ell}\mathcal{M}^{-1}_{F,r}\mathcal{M}_Sx_S\mathcal{M}^{-1}_N c_{N,\sigma}$ in the $N/F/S$ case and 
$x_X=\mathcal{M}^{-1}_X\mathcal{M}_{S1,\ell}\mathcal{M}^{-1}_{S1,r}\mathcal{M}_{F,\ell}\mathcal{M}^{-1}_{F,r}\mathcal{M}_{S2}x_{S2}\mathcal{M}^{-1}_X c_{X,\sigma}$
%EMpaper2 - missing term involving c added on
%otv3 is an additive c term missing somewhere above? 
in the $S/F/S$ case with normal contact $X$. The matrices $\mathcal{M}$ are obtained
just as the $M$ matrices below Eq.~(\ref{Xwavedown}) %otv3
from the continuity conditions,
of which now there is only at each interface.
This can be extended to the case of an
intermediate normal metal $N$ instead of a ferromagnet simply by taking the exchange field $h$ to be zero. 

The conductance is then calculated via the BTK method and Eq.~(\ref{conductance_cond}).
The analytic solution involves inverting multiple 
$8\times8$ matrices (which can be done simply using Mathematica). However, the full solution is lengthy, %otv
excessively inscrutable and can not be simplified easily. 
Despite having an analytic solution, the form of the conductance is 
still complicated due to the sheer number of plane wave combinations of 
the $\mathcal{M}$ coefficients 
that are present in each reflection amplitude. 
Therefore, 
we do an analysis similar to that %otv
 in Ref.~\onlinecite{spinsplitpaper} for the $N/F/S$ system by considering some of the possible relevant plane wave combinations to extract periodic
behavior. 
In that work we found
that the reflection amplitudes have 
a periodicity of $2\pi/h$ (in dimensionless units) on the thickness of the $F$ layer. This lead to a periodicity of the conductance peak 
position of $\pi/h$,
with the subgap peak conductance oscillating between the zero bias and critical bias for increasing thickness of the $F$ layer.
However, there is another plane wave combination we should consider which describes the SGS for $h=0$.

Below, and in the rest of the paper, we denote dimensionless lengths 
by capital letters, e.g. $D_S\equiv k_{FS}d_S$. %otv3 not earlier? EMnov - only lowercase y, \ell, and \xi
Consider first an $N^\prime/N/S$ system (for example, a normal metal contact $N^\prime$ 
coupled to an $N/S$ bilayer). If there is an interfacial scattering barrier %otv
at the $N^\prime/N$ interface, 
it is possible for the Andreev reflected holes from the $N/S$ interface to interfere with the reflections at the $N^\prime/N$ interface. 
%there is a 
%conductance peak near the critical bias due to Andreev reflections~\cite{btk,zv2}. 
%EMpaper2 - above line does not fit.
%otv2 I removed ref 50
We may look for resonance effects in the $N^\prime/N/S$ system by
examining the plane wave combination $e^{ik^+_{N}D_N}e^{-ik^-_{N}D_N}$ at the critical bias $\epsilon=\Delta_0$. 
The wavenumber in the normal metal is then
\begin{equation}
k^{\pm}_{N}=\left[1\pm{\Delta_0}\right]^{1/2}\approx1\pm\Delta_0/2
\label{kNeq}
\end{equation}
The combination is in resonance when $e^{i\Delta_0 D_N}=e^{2\pi i n}$ where $n$ is the integer of the harmonic resonance. 
Thus, the resonance in the amplitudes is expected to occur for  
\begin{equation}
 D_N=\frac{2\pi}{\Delta_0} n = \pi^2 n \Xi_0 
\label{resonanceFS}
\end{equation}
where the normalized pair potential $\Delta_0$ is related to~\cite{Kittel,BCS} 
the (dimensionless) coherence length $\Xi_0$ by 
$\Delta_0={2}/{(\pi\Xi_0)}$. 
The conductance is proportional 
to the absolute square of the amplitudes, thus the periodicity in the conductance peak resonance occurring at the critical bias should be 
\begin{equation}
\lambda_n=\frac{\pi^2}{2} n \Xi_0 
\label{lambda}
\end{equation}
In Sec.~\ref{Results_NSC_spinvalve} we will discuss
(see Fig.~\ref{NSCfig1_JJ}) %otv
 the calculated conductance for varying thicknesses $D_N=\lambda_n$. 
%otv What we see is that 
We will see that the $\lambda_n$ periodicity describes the formation of 
{\it new} peaks at the critical bias, shifting the previous $n$ numbered peak into the subgap. 
%otv? as we discuss in greater detail therein. 
This resonance is the result of multiple Andreev reflections, 
where an electron/hole is Andreev reflected off the $S$ layer and is again reflected at the $N^\prime/N$ interface. 
The integer $n$ is the harmonic of this resonance effect. For $h\neq0$, there is an additional oscillatory behavior due to the spin-split
effect described in Ref.~\onlinecite{spinsplitpaper}, which we discuss in Sec.~\ref{Results_NSC_spinvalve} as well (see Fig.~\ref{NSCfig2_JJ}). %EMnov made parenthetical
%otv i edited above but it does repeat Sec IIIA
%EMpaper2 - it's ok, it bares repeating a few of these details

In the Josephson structure $S/N/S$, the additional $S$ layer %otv edits below
leads %otv3
 to another layer thickness dependence on the conductance. 
Andreev reflected electrons and holes
from the $N/S_2$ interface may be Andreev reflected again at the $S_1/N$ interface. 
If we again consider a normal metal contact $X$ with interfacial scattering at the $X/S_1$ contact,
the quasiparticles which transmit through the $S_1$ layer may also reflect at the $X/S_1$ contact.
The net result is a complex resonance effect that can be divided into two parts: resonance
from reflections at the $X/S_1$ interface and from refections
 at the $S_1/N$ boundary. We do not have a simple argument for the exact resonance behavior and use
a phenomenological approach. We first assume a resonance effect similar to Eq.~(\ref{lambda}). Then, we introduce a term $Q$ to take into account
the actual computed %otv
dependence of the resonance on $D_S$. We find two harmonic resonance effects on $D_N$, labeled as the even and odd harmonics:
\begin{align}
\label{ResonanceJJ}
\frac{\lambda_{n,even}}{\Xi_0}=\frac{\pi^2}{2} n, \quad n=0,2,4,... \nonumber \\
\frac{\lambda_{n,odd}}{\Xi_0}=\frac{\pi^2}{2} n - Q\left(\frac{D_{S_1}}{\Xi_0}\right), \quad  n=1,3,5,...
\end{align}
where we find that
$Q\left(D_{S_1}/\Xi_0\right)\approx 1.2 \ln\left(D_{S_1}/\Xi_0\right) + 1.94$ approximates the resonance values. 
The even terms are due to reflections at the $S_1/N$ interface and have the same form 
as Eq.~(\ref{lambda}) while the odd terms are due to
reflections at the $X/S_1$ contact interface. The odd resonance values are reduced by a term %otv
 $Q$ which depends only on the ratio $D_{S_1}/\Xi_0$. 
%At the critical bias  %otv
A new peak forms at the critical bias for $D_N=\lambda{n}$, shifting lower order
harmonic peaks %otv2left 
to lower bias, %otv2
into the subgap as $D_N$ increases. These peaks
are equally separated between even/odd pairs for constant $D_{S_1}/\Xi_0$. 
%EMpaper2 - more edits above
We study these peaks for multiple harmonics
in the $h=0$ case and the $h\neq0$ case in Sec.~\ref{Results_NSC_JJ},
see the discussion below
associated with Figs.~\ref{NSCfig3_JJ} and \ref{NSCfig4_JJ}. %otv. 

We will not consider the higher harmonics ($n\geq1$) when %otv3
using our numerical method.
This is for two reasons: first, the peak positions are much more difficult to predict as the saturated pair potential 
(or the ``effective'' coherence length $\Xi=2/{(\pi\Delta})$) is not constant 
as one varies layer thicknesses due to the proximity effect. Second, the $n=1$ harmonic occurs for very large intermediate thicknesses:
about five times the coherence length of the superconductor. The 
ballistic nanostructures we wish to study (those built by experimentalists) 
typically have a total intermediate thickness less than or on the same order as 
the coherence length. %otv which is why we study these systems in the ballistic limit. 
By introducing ferromagnets we can probe the higher harmonic peaks at lower intermediate thicknesses
due to the oscillatory behavior of the peaks. We present analytic
results in Sec.~\ref{Results_NSC_JJ} and numerical ones
in Sec.~\ref{Results_XSFS_df2} for the $S_1/F_1/N/F_2/S_2$ ferromagnetic Josephson structure.

%%%%%%%%%%%%%%%%%%%%%%
\section{Results}
\label{JJresults}

In this section we present our results on the conductance in the ferromagnetic Josephson structures ($S/F/S$ and $S/F/N/F/S$). 
%As in  previous work\cite{condpaper,spinsplitpaper} %otv
%we focus on the forward conductance at the $T\rightarrow0$ limit.  %otv below rewritten
%otv3 Eq (6) in only valid at low T and I added this earlier. I am not
% sure  where the forward limit is taken 
Preliminarily to our numerical results, we will
consider a simplified model of the $N/F/S$ structure, which can
be treated analytically, and work our way up to the $S/F/S$ structure. 
Although this model is quantitatively inaccurate,
it does help highlight the qualitative features of the subgap conductance
in a more intuitive manner. This qualitative understanding
 is very useful when describing
the fully self-consistent numerical results of the $S_1/F_1/N/F_2/S_2$ structure. We include %otv3
 a normal metal contact $X$ %otv3 repeat which simplifies the model 
as discussed in Sec.~\ref{JJqpCondSub}, in all our results on Josephson structures. 
%otv3 moved to later so that the system is $X/S/F/S$ or $X/S_1/F_1/N/F_2/S_2$. 
In our numerical results, we determine their dependence
on the  thickness  of the $F_2$ layer as well as on $\phi$, 
the misalignment angle of the $F$ layer magnetizations. 
The thickness dependence will be described 
in relation to the analytic results.

Our results are parameterized by the layer thicknesses and the coherence length of the superconductor. As mentioned above, %otv3
length scales are all
normalized %otv3 them
to $k_{FS}$, the Fermi wavevector of the superconductor. %otv some rewrite
All energies are understood to be normalized to $E_{FS}$ %otv3 notation inconsistency fixed
except the dimensionless bias $E$, which is is normalized to
 the bulk pair potential $\Delta_0$. The conductance is in units of $2\pi e/\hbar$. The interfacial
scattering barriers $H_B$ are normalized by $v_F$. In each figure below, 
we take the scattering at the left-side contact interfaces ($X/S$, $N/F$, and $N^\prime/N$) to the intermediate value, $H_B=0.5$. %otv
This barrier enhances
many of the subgap conductance features by making the peaks sharper and we
believe it better represents a realistic experimental situation. In the numerical
results, we also consider ideal %otv3 clean 
contact interfaces for comparison.
%EMpaper2 - we now also have H_B0 = 0$
In the analytic results we assume no interfacial scattering at the $N/S$ and $F/S$ interface, for simplicity, 
but in the numerical calculations we consider both zero and non-zero interfacial scattering at the intermediate $F/N$ interfaces. 
We take the normalized ferromagnetic exchange field to be $h=0.145$, %otv3 repetitive normalized by $E_F$, 
and the dimensionless coherence length to be $\Xi_0=115$ for each ferromagnetic and superconducting layer respectively. These values 
have been found to be suitable to describe experimental samples %otv3
 using cobalt and niobium~\cite{alejandro}. 

\subsection{N$^\prime$/N/S and N/F/S analytic results} 
\label{Results_NSC_spinvalve}

%otv light edits below
We start our discussion with the analytic 
results (see Sec. \ref{JJanalyticSub}) for $N/F/S$ and $N^\prime/N/S$ structures. 
In Fig.~\ref{NSCfig1_JJ} we plot the conductance of an
 $N^\prime/N/S$ multilayer for
$N$ layer thicknesses 
such that $D_N=\lambda_n$ and a single barrier $H_B=0.5$ at the $N^\prime/N$ interface. This interfacial scattering is representative
of an imperfect metallic contact interface, but it is still far from the tunneling limit. 
The left- and right-most layers ($N^\prime$ and $S$) are  infinite in thickness in the analytic approximation. 
The thicknesses $\lambda_n$ are the resonance values found in Eq.~(\ref{lambda}) and represent
the interference of the Andreev reflected electrons and holes with those reflected at the $N^\prime/N$ interface. The case $D_N=0$ is equivalent
to the $N/S$ system studied by BTK. %otv Z undefined?for $Z=0.5$~\cite{btk}. 
As seen there, the effect of the barrier decreases the conductance in the subgap region
(and at the high bias limit) without decreasing the conductance at the critical bias (CB) $eV=\Delta_{0}$. 
This leads to a sharp peak in the conductance. As the thickness $D_N$ 
increases, the critical bias peak shifts into the subgap region and a new peak is formed at the CB when $D_N$ reaches a resonance value. Increasing $D_N$ 
further, %otv3 below rewritten
other peaks form, shifting the previous peaks further towards zero bias, %and so on 
for each resonance value. We see that the peaks are evenly spaced
for each thickness plotted. Also, we see an additional oscillatory behavior in the conductance just above the critical bias. 
This oscillatory pattern decays at the same rate as in the $N/S$ case ($D_N=0$) towards the normal conductance. The frequency of the oscillations is proportional to the
harmonic $n$ of the $N$ layer thickness resonance. The thicknesses of the intermediate layer depicted in Fig.~\ref{NSCfig1_JJ} are quite large, about five times 
the coherence length of the superconductor for $\lambda_1$ and ten times for $\lambda_2$. This makes the results less relevant to the experimental %otv3
nanoscale heterostructures
that we have in mind, %otv3
where the intermediate layer thicknesses are on the order of the coherence length or less. However, this analytic calculation provides
an excellent illustration of the subgap peak structure. We will see that this structure plays a prominent role in ferromagnetic Josephson structures $S/F/S$. 
%otv3 not needed? Below we will further probe this structure by replacing the intermediate normal metal with a ferromagnet.

\begin{figure}[t]
\includegraphics[angle=-90,width=0.45\textwidth] {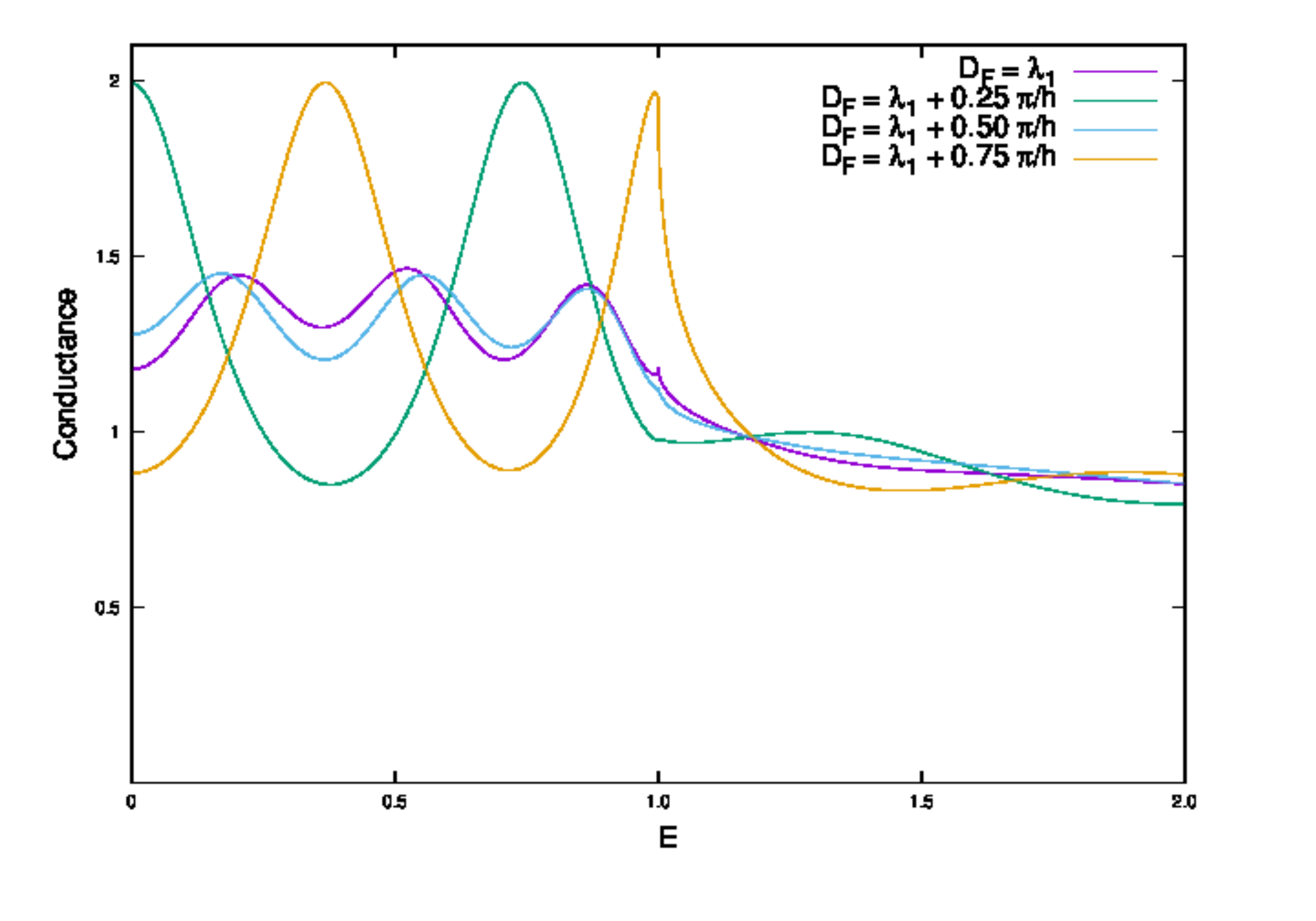} 
\vspace*{-0.2cm}
\caption{Conductance (G) vs. Bias (E) in the $N/F/S$ structure for varying $D_F$.  
$\lambda_1 = ({\pi^2}/{2})\Xi_0\approx568$ is the first harmonic resonance and ${\pi}/{h}\approx22$. The figure is plotted for one full oscillation
of the thickness dependence.
We assume a single barrier at $N/F$ interface $H_B=0.5$. Analytic approximation. 
}
\label{NSCfig2_JJ}
\end{figure}

In Ref.~\onlinecite{spinsplitpaper} %otv minor edits below
we discussed the spin-split conductance for the superconducting spin valve ($F/N/F/S$). There we also used %otv3
a similar analytic approximation for the $N/F/S$ model, although only for 
small thicknesses $D_F$. We found  that the peak conductance %otv3
oscillates between the critical bias and near zero bias with increasing thickness. 
This is due to the spin-split conductance: the conductance features differ 
for incoming spin up and spin down electrons. From our analysis we found the wavelength of the oscillations to be $\pi/h$. For these small thicknesses, there is only
one resonance peak, attributable to the $n=0$ harmonic. 
In Fig.~\ref{NSCfig2_JJ} we plot the conductance for the $N/F/S$ system, but for $D_F$ values close to the $\lambda_1$ 
resonance thickness. The periodicity of the spin-split conductance peak is significantly smaller ($\pi/h\approx22$) than the resonance thickness ($\lambda_1\approx568$ as mentioned above). %otv 
We plot in Fig.~\ref{NSCfig2_JJ} one full period of the spin split oscillation. 
We see that the conductance oscillates between two different two-peak states: one with the peaks located at the CB and %otv2left %otv? %EMpaper2 - down? left or right
at lower bias, %otv2 OK now?
near the middle of the subgap, and one with the peaks located from near zero bias 
and right to the middle of the subgap. As the thickness increases, the set of 
two peaks oscillates between %otv3 set is singular
low biases and the critical bias, as is the case\cite{spinsplitpaper}
 for  small $F$ thicknesses. Between these states, each peak splits into two, which corresponds to
the subgap peaks found in the small $D_F$ case, and features a cusp peak near the CB. These subgap peaks are a split of the two resonance peaks for the first ($n=1$) harmonic.
Because this oscillatory behavior effectively shifts all resonance peaks further into the subgap, it is possible to have multiple subgap peaks for thicknesses less than the first harmonic
resonance thickness as the higher order resonance peak will shift from the CB into the subgap region. 
However, the thickness must still be much larger than the coherence length to see this effect in $N/F/S$ systems.

\begin{figure}[ht]
\includegraphics[angle=-90,width=0.45\textwidth] {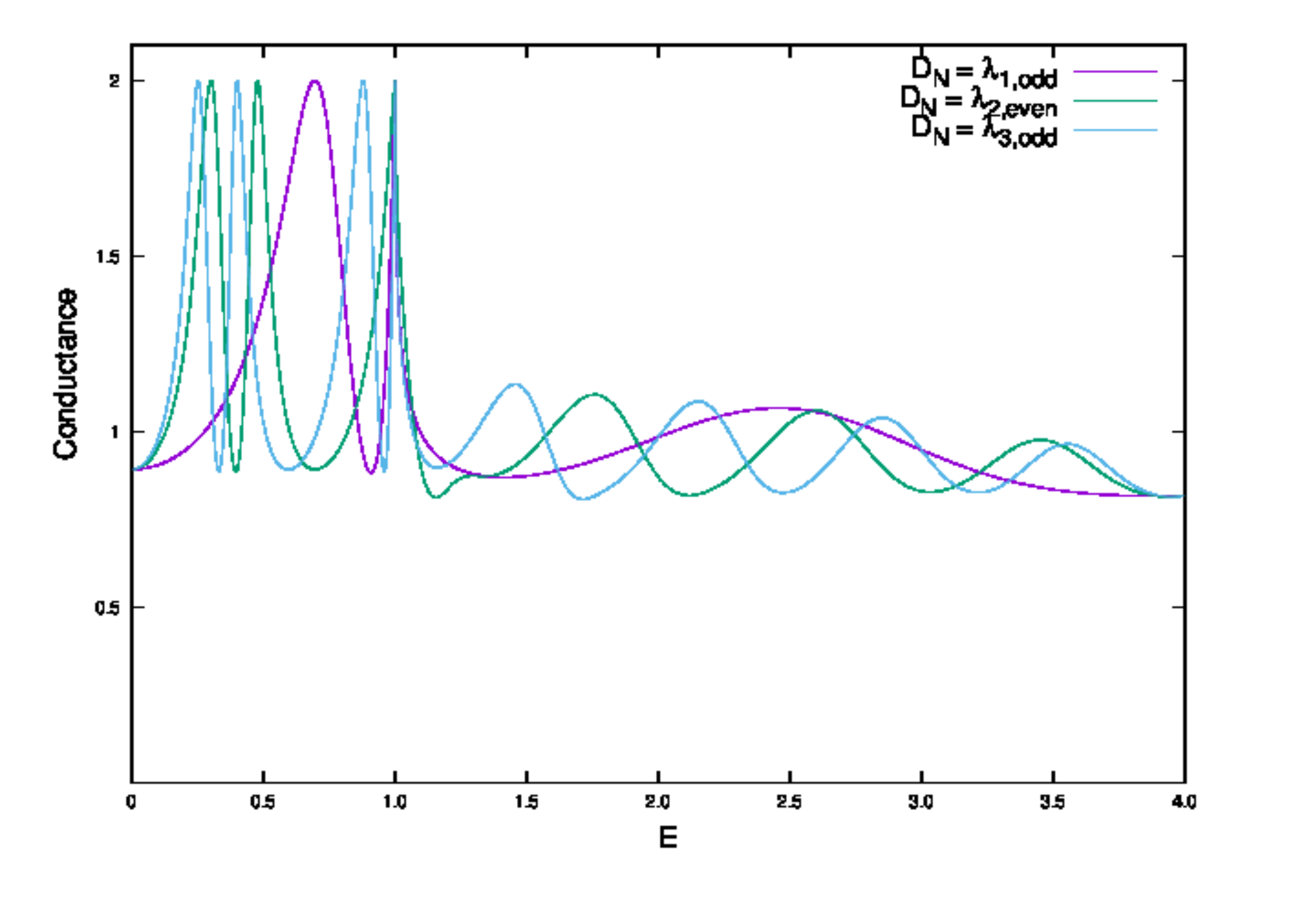} 
\vspace*{-0.3cm}
\caption{Conductance (G) vs. Bias (E) in the $S_1/N/S_2$ structure for $D_{S1}=180$ varying $D_N$. 
See Eq.~(\ref{ResonanceJJ}) for thicknesses $\lambda_n$
and the relevant discussion. The ``odd'' peaks (see text) %otv3
are shifted by a constant in their resonance values. 
We assume a single barrier at the $X/S_1$ contact $H_B=0.5$. Analytic approximation. 
}
\label{NSCfig3_JJ}
\end{figure}

\subsection{S/N/S and S/F/S analytic results}
\label{Results_NSC_JJ}

We now turn our attention to analytic 
results for %otv
$S_1/N/S_2$ and $S_1/F/S_2$ Josephson structures. In Fig.~\ref{NSCfig3_JJ} we plot the conductance for the $S_1/N/S_2$
structure with $D_{S1}=180$ and %otv3
 resonance values of the $N$ layer thickness (Eq.~\ref{ResonanceJJ}). %otvWe also
%include a normal metal contact $X$ with nonzero 
The interfacial scattering at the $X/S_1$ contact is $H_B=0.5$. %otv3
As discussed in Sec.~\ref{JJanalyticSub} the Josephson structure has two sets of %otv3 
resonance
values on $D_N$: the ``even'' and ``odd'' resonances. The even resonances are the same as for the $N^\prime/N/S$ case, but the odd resonances have an extra
term %otvfactor 
$Q\left(D_{S1}/\Xi_0\right)$ (see Eq.~\ref{ResonanceJJ}) 
that decreases the resonance thickness for the odd $n$ harmonics (from $\lambda_1\approx568$ to $282$). 
This split in resonances
is due to the difference in the reflections at the $S_1/N$ and the $X/S_1$ interfaces.
The exact form of the additional %otv3
term was not determined, but from analyzing results such as those shown in Fig.~\ref{NSCfig3_JJ} 
we were able to estimate %otv
 the value of $Q$ as in Eq.~(\ref{ResonanceJJ}).  %otv
We plot a range of thicknesses which include the first two odd resonances ($n=1,3$) as well as the $n=2$ even resonance. 
As $D_N$ increases, we see the same shift %otv
of the critical bias peaks into the
subgap region. However, due to the dual resonance structures, these peaks are not evenly spaced. 
Furthermore, the oscillations above the gap are not in phase and the frequency is not directly
proportional to the harmonic $n$ for the odd resonances.

\begin{figure}[!t]
\includegraphics[angle=-90,width=0.45\textwidth] {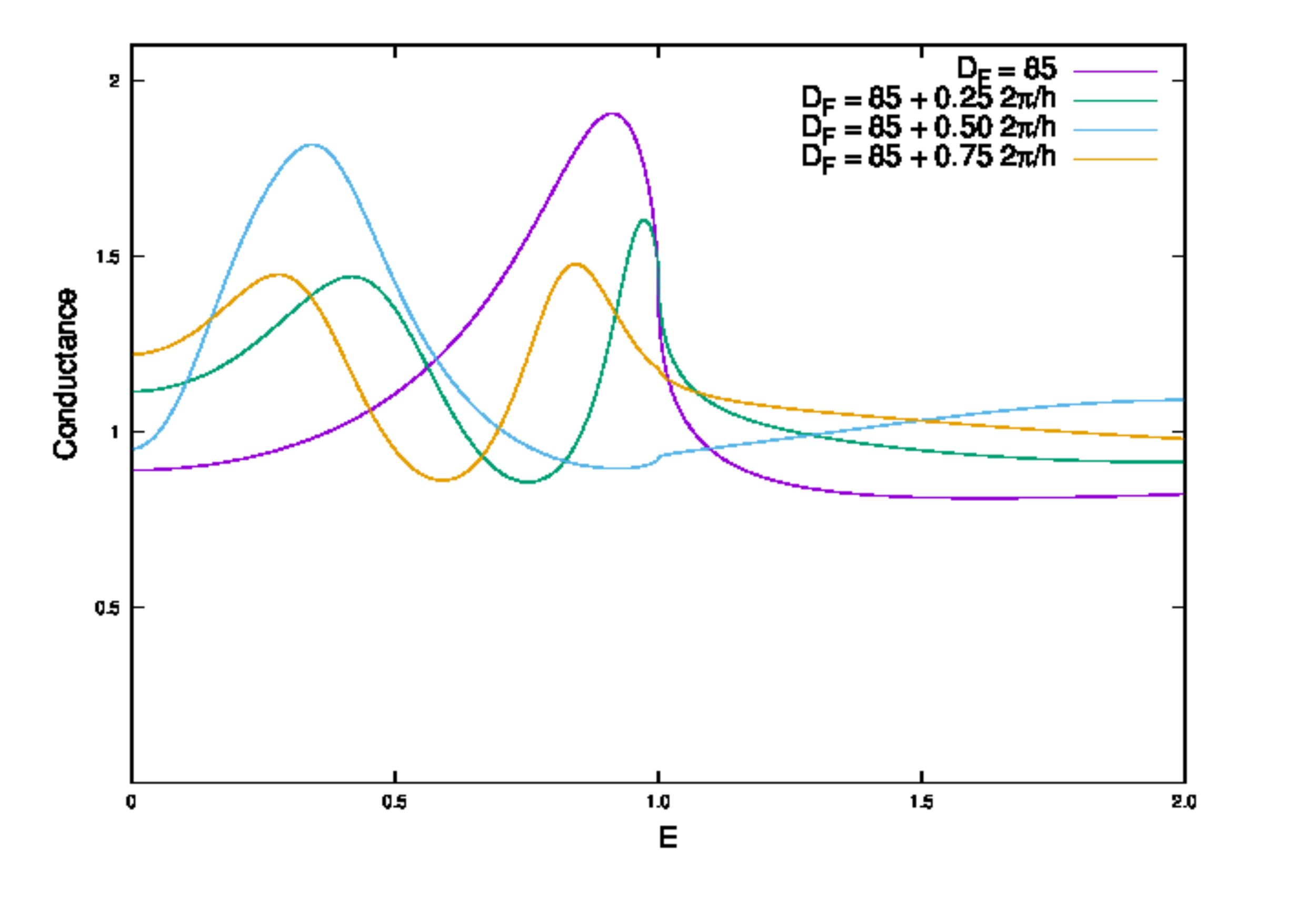} 
\vspace*{-0.3cm}
\caption{Conductance (G) vs. Bias (E) in the $S_1/F/S_2$ structure for $D_{S1}=180$ and %otv3
varying $D_F$. 
We plot the conductance for one full oscillation of the thickness periodicity $2\pi/h\approx43$. 
We assume a single barrier at the $X/S_1$ contact $H_B=0.5$. Analytic approximation. 
}
\label{NSCfig4_JJ}
\end{figure}

In Fig.~\ref{NSCfig4_JJ} we plot the conductance for the $S_1/F/S_2$ structure in our analytic approximation. %otv repeat which includes an imperfect contact $X$. 
We have previously established\cite{spinsplitpaper}
that the conductance peak is oscillatory between low biases and the CB for varying $D_F$ thicknesses in $N/F/S$ structures. 
In the previous subsection, we have shown that %otv
this extends to all resonance peaks.
We now do the same analysis for the ferromagnetic Josephson structure. 
We see an oscillatory behavior due to the spin-split conductance
that is similar to that in %otv3
the $N/F/S$ case, except that 
the total periodicity is now $2\pi/h\approx43$.
We have set the minimum thickness %otv
to be $D_F=85$ which is less than the first $n=1$ resonance value ($\lambda_{1,odd}\approx282$). 
This is for two reasons: %otv
 First, this value is the minimum
total thickness of the intermediate layers (between $S_1$ and $S_2$) in our numerical calculations on the $S_1/F_1/N/F_2/S_2$ ferromagnetic Josephson structure. Second,
we wish to show how the oscillations of the resonance peaks can shift a higher order harmonic peak into the subgap region, allowing for multiple subgap peaks.
Indeed, in Fig.~\ref{NSCfig4_JJ} we see a single conductance peak at $D_F=85$. As $D_F$ increases, it splits into two subgap peaks (with one being very near the CB). Then, the two
peaks reform at $D_F=85+\pi/h$ into a single subgap peak. Increasing $D_F$ %otv
further, this peak splits into two subgap peaks with one being at very low biases. This is quite different
from what was %otv3
 found in the $N/F/S$ structure. Not only is the overall periodicity of the behavior doubled, but here there may exist multiple, distinct subgap peaks instead of a single
peak and a cusp at the critical bias. This occurs for realistic thicknesses of the intermediate layers, at least in our analytic approximation. In the next subsection,
we analyze the fully self-consistent $S_1/F_1/N/F_2/S_2$ structure.

\subsection{Self-consistent, numerical results for the %otv
S$_1$/F$_1$/N/F$_2$/S$_2$ conductance: F$_2$ layer thickness dependence} 
\label{Results_XSFS_df2}

Through our approximate analytic study, we have found that there are 
two sources of resonance in the $S/F/S$ Josephson structure that give rise to %otv
 conductance peaks in the subgap region. 
Furthermore, these peaks are oscillatory with increasing thickness of the $F$ layer. We now discuss the numerical results of the ferromagnetic Josephson structure $S_1/F_1/N/F_2/S_2$. 
We include a thin normal metal contact $X$ which allows us to simplify our methods
as explained %otv3
 in Sec.~\ref{JJqpCondSub}, the  structure being then $X/1/F_1/N/F_2/S_2$.
%otv3 moved from earlier
We will consider both zero and nonzero
interfacial scattering at the $X/S_1$ contact, and we compare the dependence of the conductance on the $F/N$ and $F/S$ interfacial scattering.
In numerical calculations, 
the pair potential within each superconductor is a function of position within the multilayer, as determined by our 
self-consistent method. Furthermore, %otv3 each layer  has a finite thickness. 
in the numerical calculation, all layers are finite in width. We keep all
layer thicknesses constant except the $F_2$ layer: $D_{S1}=D_{S2}=180$, 
$D_{F1}=30$ and $D_N=40$. The normal metal contact thickness is $D_X=5$. 
We set again $h=0.145$ and $\Xi_0=115$. %otv3
%otv3 repeated The metallic contact is thin enough not to affect the proximity effect. 
Our results focus on
the quasiparticle current, and do not reflect the zero bias current due to the Josephson effect. Therefore, in interpreting our results, it should be noted
that the ultra-low bias conductance may be inaccessible in experiment, even with a hysteresis current from a shunted Josephson circuit, due to the Josephson current.

\begin{figure*}[!pth]
\centering
\vspace*{-0.5cm}
\begin{subfigure}[b]{\textwidth}
\includegraphics[width=0.48\textwidth] {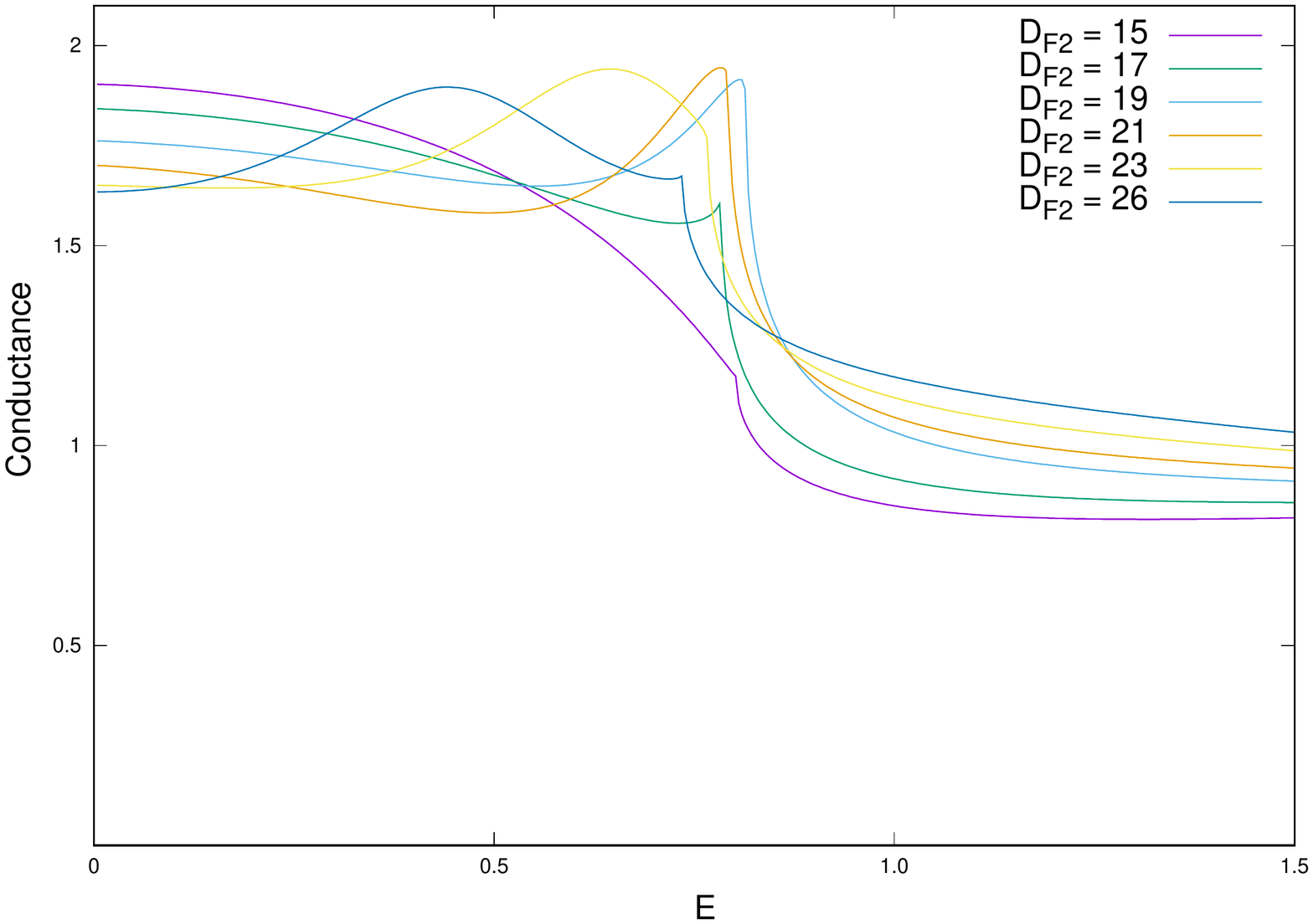}
\vspace*{-0.1cm}
\caption{Barrier at the $F/N$ interfaces $H_{B,0}=H_{B,1}=H_{B,4}=0$, $H_{B,2}=H_{B,3}=0.5$.
}
\label{XSFSfig2b_noX}
\end{subfigure}
\begin{subfigure}[b]{\textwidth}
\includegraphics[width=0.48\textwidth] {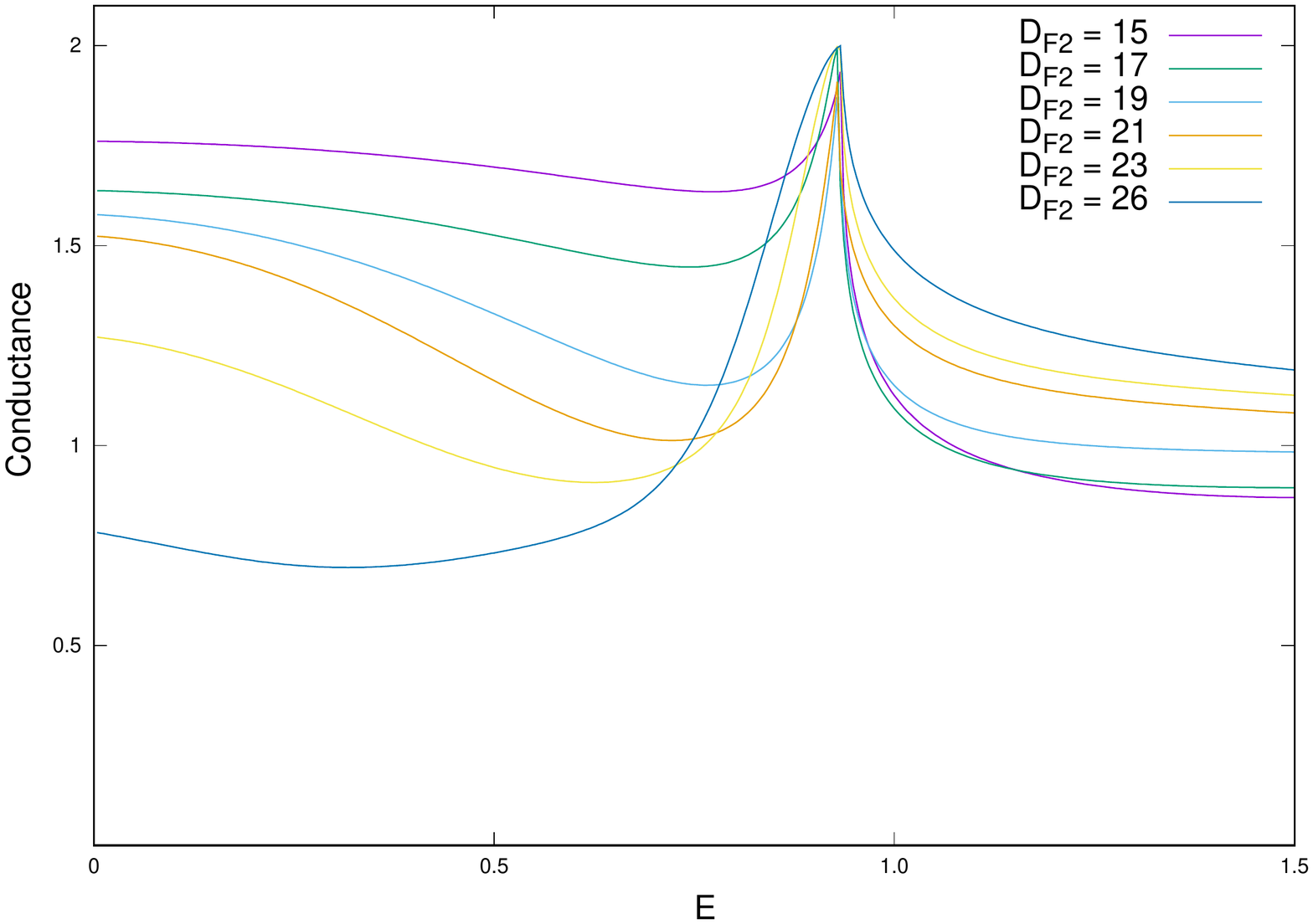}
\vspace*{-0.1cm}
\caption{Barrier at the $F/S$ interfaces $H_{B,0}=H_{B,2}=H_{B,3}=0$, $H_{B,1}=H_{B,4}=0.5$.
}
\label{XSFSfig2c_noX}
\end{subfigure}
\caption{Numerical results for %otv3
the conductance (G) vs. Bias (E) in the $S_1/F_1/N/F_2/S_2$ structure for varying $D_{F2}$ and $\phi=0$ with %otv 
transparent $X/S_1$ interface and other %otv3
interfacial scattering values as indicated.}
\label{XSFSfig2_JJ_noX}
\end{figure*}

\begin{figure*}[!pth]
\centering
\vspace*{-0.5cm}
\begin{subfigure}[b]{\textwidth}
\includegraphics[angle=-90,width=0.48\textwidth] {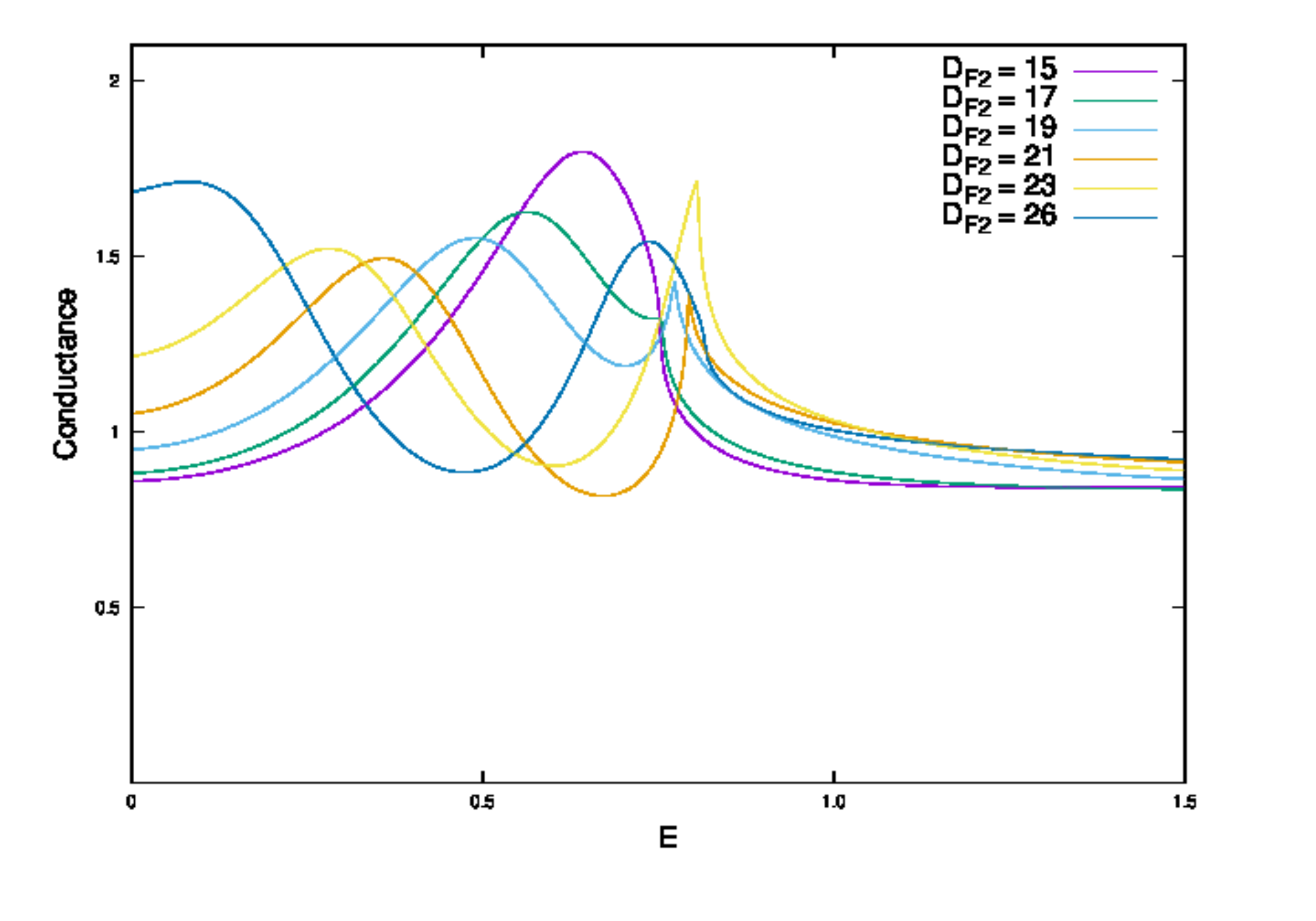}
\vspace*{-0.1cm}
\caption{Single barrier at the $X/S_1$ contact $H_{B,1}=H_{B,2}=H_{B,3}=H_{B,4}=0$, $H_{B,0}=0.5$.
}
\label{XSFSfig2a}
\end{subfigure}
\begin{subfigure}[b]{\textwidth}
\includegraphics[angle=-90,width=0.48\textwidth] {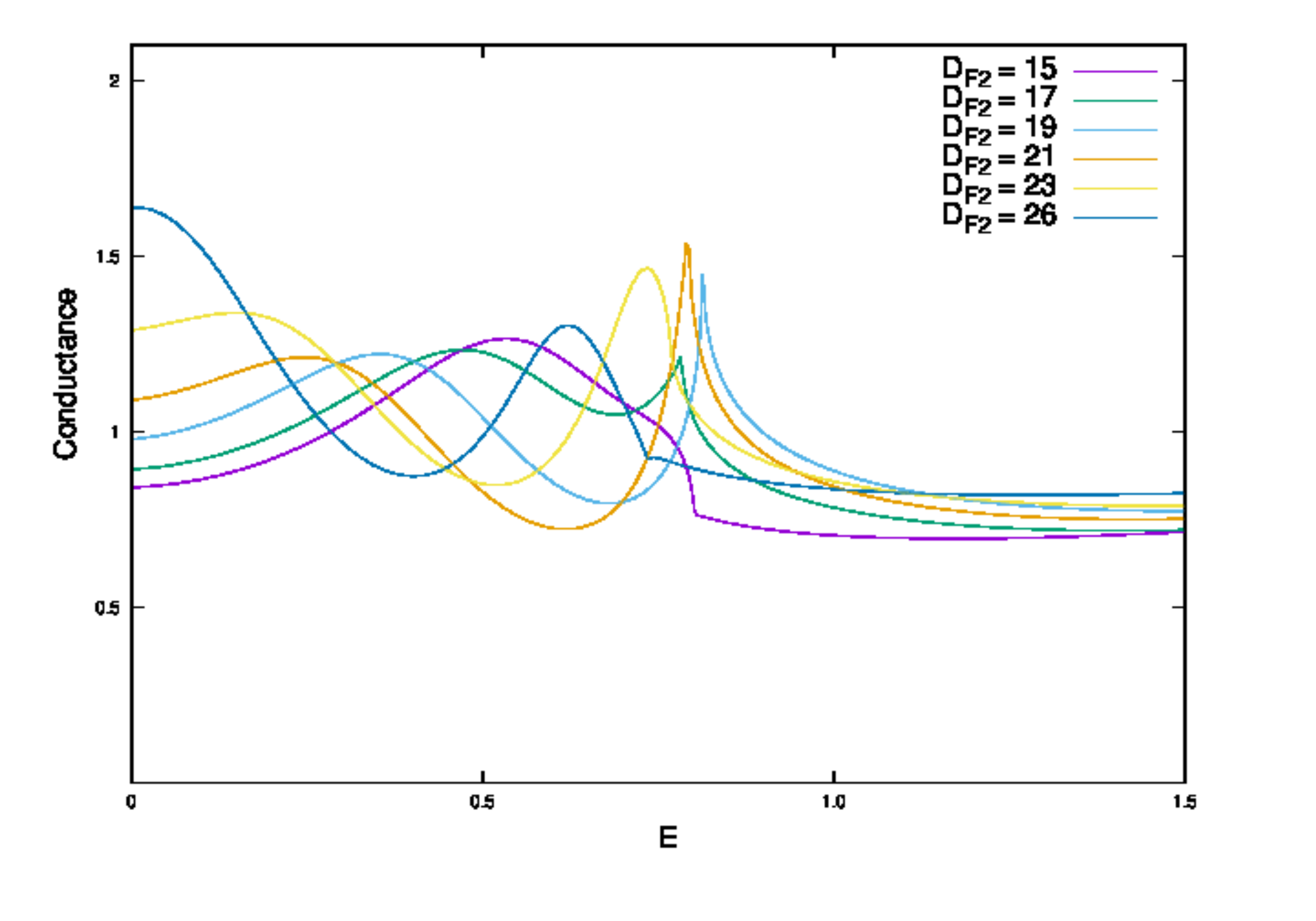}
\vspace*{-0.1cm}
\caption{Barrier at the $X/S_1$ and $F/N$ interfaces $H_{B,1}=H_{B,4}=0$, $H_{B,0}=H_{B,2}=H_{B,3}=0.5$.
}
\label{XSFSfig2b}
\end{subfigure}
\begin{subfigure}[b]{\textwidth}
\includegraphics[width=0.48\textwidth] {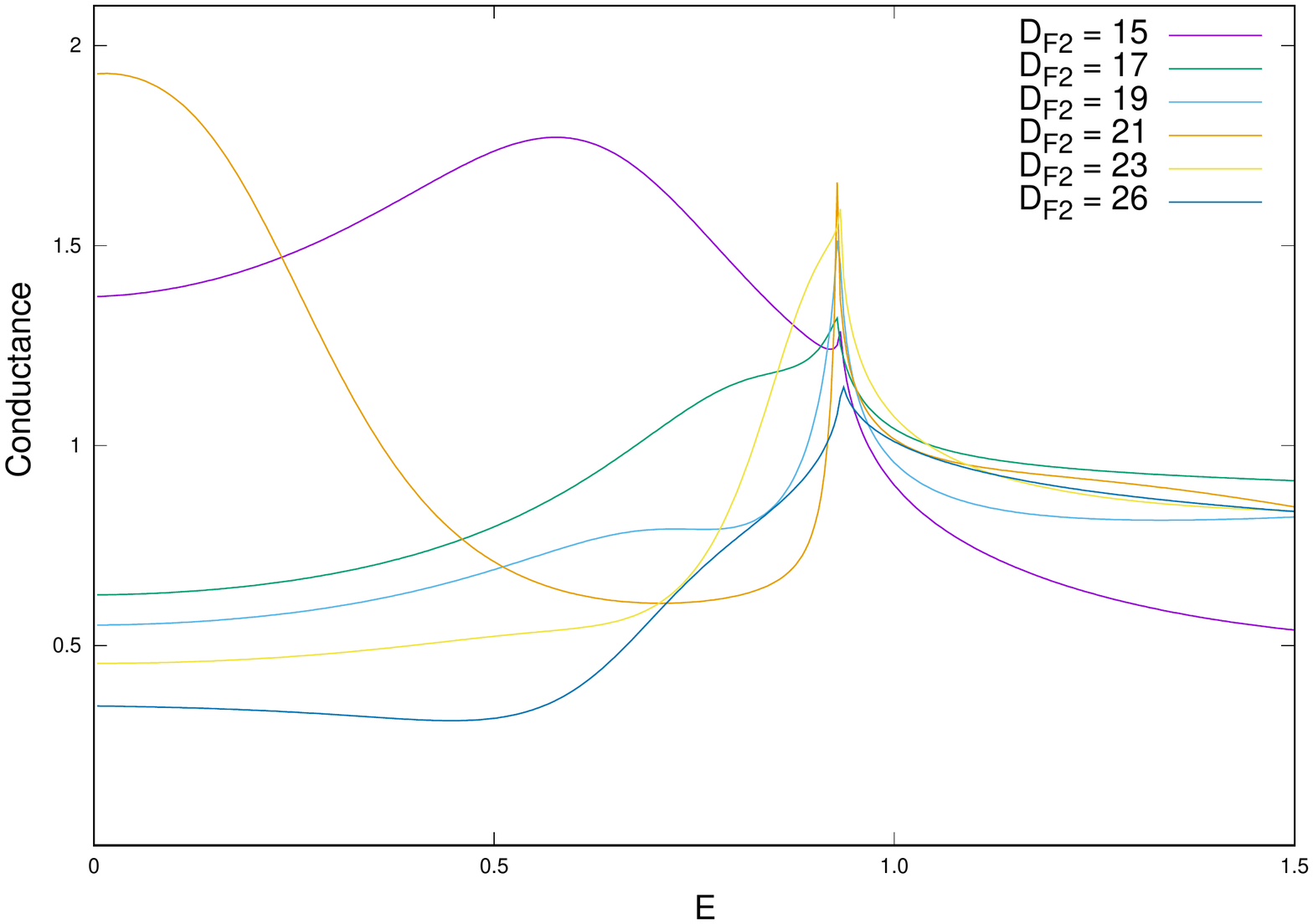}
\vspace*{-0.1cm}
\caption{Barrier at the $X/S_1$ and $F/S$ interfaces $H_{B,2}=H_{B,3}=0$, $H_{B,0}=H_{B,1}=H_{B,4}=0.5$.
}
\label{XSFSfig2c}
\end{subfigure}
\caption{Numerical results for the %otv3
conductance (G) vs. Bias (E) in the $S_1/F_1/N/F_2/S_2$ structure for varying $D_{F2}$ and $\phi=0$. The interfacial scattering at $X/S$ contact $H_{B,0}=0.5$ and the others are as indicated in each panel.
} %otv
\label{XSFSfig2_JJ}
\end{figure*}

In this subsection, we determine the dependence of the conductance 
features on $D_{F2}$. The magnetization of the ferromagnetic
layers are assumed to be parallel ($\phi=0$). In Sec.~\ref{Results_XSFS_phi} 
we will consider the dependence of $G$ on the
misalignment angle $\phi$ at fixed $D_{F2}$. %otv moved and changed
We consider several sets of interfacial scattering strengths.
The interfacial scattering strengths $H_{B,i}$ are indexed from the 
far left $X/S_1$ to the right $F_2/S_2$ %otv3 no X in Fig 1(see Fig.\ref{JJfig0})
 starting from zero
($0\le i\le4$) thus
the $X/S_1$ barrier strength is $H_{B,0}$, the $F/N$ interfacial barrier strengths are $H_{B,2}$ and $H_{B,3}$, and the $F/S$ barrier strengths are $H_{B,1}$ and $H_{B,4}$. The barrier strengths are either zero or $0.5$. 
We do not consider the situation with all transparent interfaces, i.e. $H_{B,i}=0$, as this leads to minimal subgap structure and it would be quite unrealistic
experimentally.  %otv

In Fig.~\ref{XSFSfig2_JJ_noX} we study the conductance dependence for a transparent $X/S_1$ interface ($H_{B,0}=0$). It has two subfigures: 
(a) with $F/N$ interfacial scattering, $H_{B,2}=H_{B,3}=0.5$,
and (b) with $F/S$ interfacial scattering, $H_{B,1}=H_{B,4}=0.5$. 
Conversely, in
 Fig.~\ref{XSFSfig2_JJ} we
consider the effect of interfacial scattering at the $X/S_1$ contact interface, 
$H_{B,0}=0.5$, in three subfigures (a) no additional barriers, (b) $F/N$ interfacial scattering,
and (c) $F/S$ interfacial scattering.
In both of these
figures  we plot the dimensionless conductance for $\phi=0$ and varying $D_{F2}$.
%In Fig.~\ref{XSFSfig2_JJ_noX} we have $H_{B,0}=0$ and in Fig.~\ref{XSFSfig2_JJ} 
%a nonzero contact barrier . 
In Fig.~\ref{XSFSfig2b_noX} we see in the subgap region a single peak structure,
where the peak of the conductance moves from low bias to the critical bias over the range of $D_{F2}$ shown. This single peak structure is similar to %otv?
 that studied in
Ref.~\onlinecite{spinsplitpaper} in which a subgap peak was shown to oscillate in bias position with the ferromagnetic layer thickness in $F/N/F/S$ spin valve structures which
also had interfacial scattering at the $F/N$ interfaces. 
The periodicity of the oscillations was shown to be $\pi/h$. When the peak is located  between zero bias and the critical bias, there is a cusp feature 
in the critical bias conductance.
In subfigure Fig.~\ref{XSFSfig2c_noX}, where %otv
 we plot the conductance for nonzero $F/S$ scattering,
we see a very different phenomenon:  there is now a subgap minimum
as opposed to a subgap peak. There is a marked
peak at the critical bias. The subgap minimum is also dependent on $D_{F2}$. We see that the subgap minimum goes from being %otv
near the critical bias at $D_{F2}=15$ to being near zero bias at $D_{F2}=26$. This thickness dependence is different from the oscillatory peak structure in 
Fig.~\ref{XSFSfig2b_noX}; the periodicity of the minima is not $\pi/h$. This minimum, or dip structure is not unprecedented. In Ref.~\onlinecite{OBTK}, %OBTK
a subgap peak structure is found in the {\it resistance} %otv %EMpaper2 - some fixes
when considering nonzero interfacial scattering at the $S/N$ interfaces in $S/N/S$ Josephson junctions.
This translates to dips in the conductance.

In Fig.~\ref{XSFSfig2_JJ} we plot the conductance with a nonzero barrier at the $X/S_1$ interface $H_{B,0}=0.5$. The results are more 
qualitatively similar to the analytic results presented in Sec.~\ref{Results_NSC_JJ}.
From $D_{F2}=15$ to $D_{F2}=26$ the behavior of $G$
with bias changes
 from a single peak near the CB to a two-peak structure, one at low bias and one just below the critical bias. As
$D_{F2}$ increases from $15$, we see the single peak shift into the 
subgap region %otv
until a second peak forms at the CB at around $D_{F2}=21$. The thickness difference between
$D_{F2}=15$ and $D_{F2}=26$ is about $\pi/2h$, which is one quarter of the total oscillatory pattern (see Fig.~\ref{NSCfig4_JJ} for comparison). This means the periodicity is doubled
by the presence of the $X/S_1$ layer. This is due to the reflections at 
the $X/S_1$ interface which form a second resonance effect as described in Sec.~\ref{Results_NSC_JJ}.
Comparing Figs.~\ref{XSFSfig2a} and \ref{XSFSfig2b}, we see the effect that 
the $F/N$ barriers have on the conductance. %otv
We observe that the phase of the oscillatory spin-split behavior
shifts slightly and the conductance decreases. The subgap peak structure is 
{\it not} enhanced by the barriers, but instead the conductance is decreased in all peak values. 
The biggest change occurs when we shift focus to the $F/S$ barriers in Fig.~\ref{XSFSfig2c} where the subgap sctructure is much more complicated: there is no subgap peak,
except at $D_{F2}=15$ and $D_{F2}=21$, but there is a noticeable inflection point in the subgap at the other thicknesses. We ascribe this to the combined effect of the peak structure
we see in Fig.~\ref{XSFSfig2a} and the dip, or minima structure we see in Fig.~\ref{XSFSfig2c_noX}. The presence of the $F/S$ barriers forms dips while the $X/S_1$ barrier provides a
peak resonance in the MAR. These two effects 
 do not share the same periodicity with the thickness, which complicates the overall effect. From this we conclude
that the $F/S$ barriers have the most impact on the subgap structure. The $X/S_1$ barrier is also important as it leads to a resonance effect in the MAR which can lead to a 
second conductance peak or a complex inflection structure within the subgap, depending on what other barriers are in play.

\subsection{Self-consistent, numerical S$_1$/F$_1$/N/F$_1$/S$_1$ conductance: angular dependence}
\label{Results_XSFS_phi}

\begin{figure*}[!pth]
\centering
\vspace*{-0.5cm}
\begin{subfigure}[b]{\textwidth}
\includegraphics[width=0.48\textwidth] {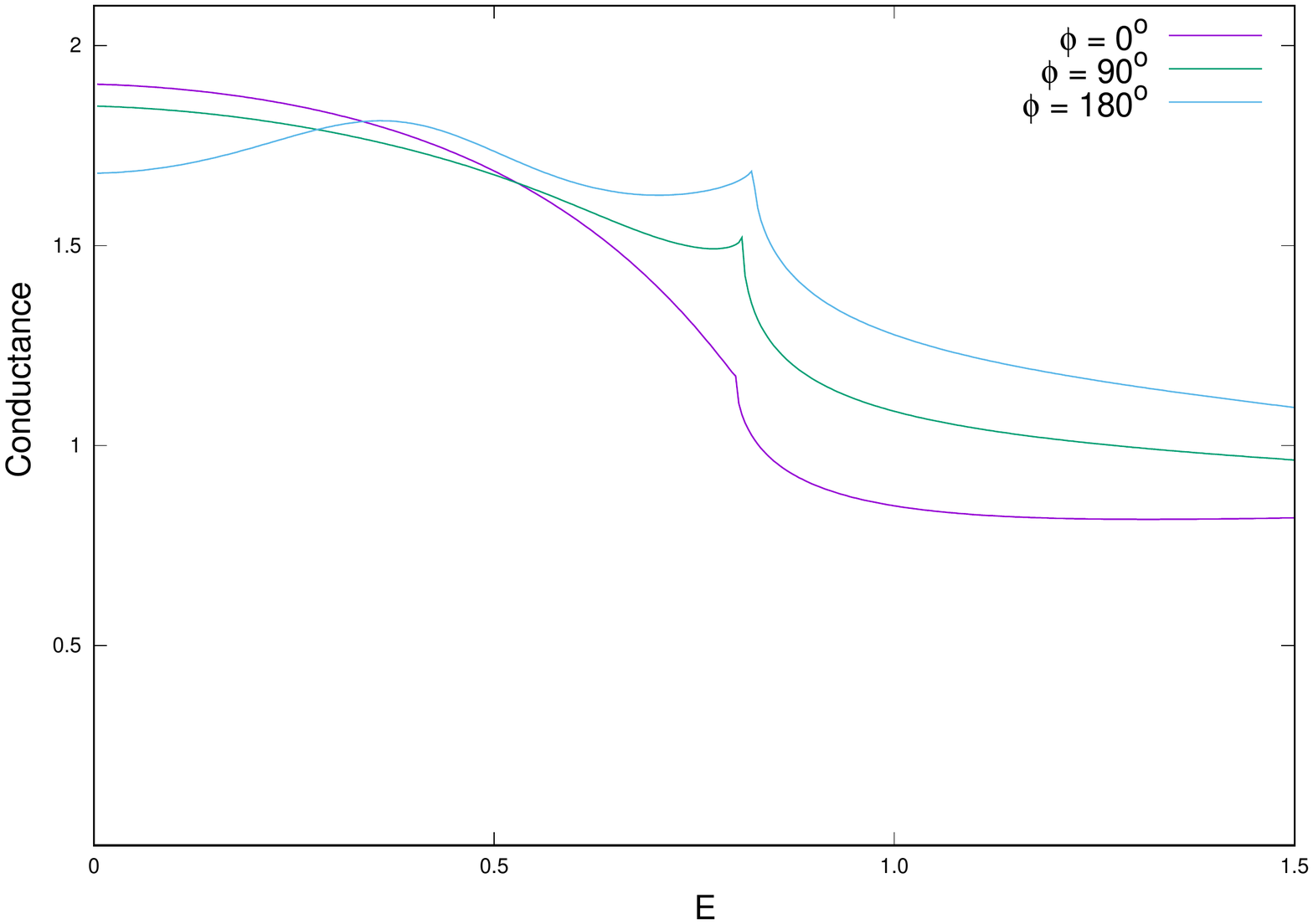}
\vspace*{-0.1cm}
\caption{Barrier at the $F/N$ interfaces $H_{B,0}=H_{B,1}=H_{B,4}=0$, $H_{B,2}=H_{B,3}=0.5$.
}
\label{XSFSfig3b_noX}
\end{subfigure}
\begin{subfigure}[b]{\textwidth}
\includegraphics[width=0.48\textwidth] {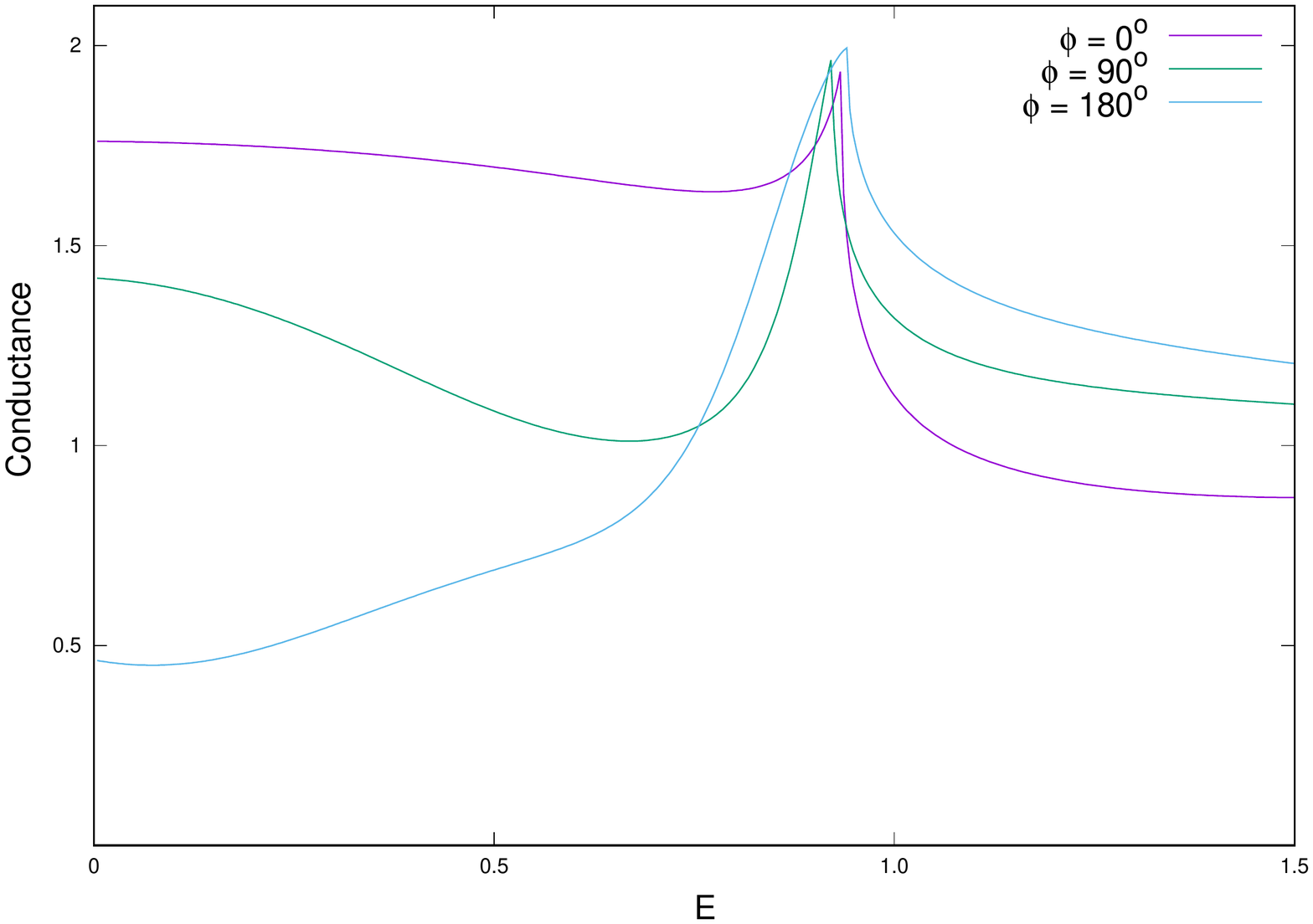}
\vspace*{-0.1cm}
\caption{Barrier at the $F/S$ interfaces $H_{B,0}=H_{B,2}=H_{B,3}=0$, $H_{B,1}=H_{B,4}=0.5$.
}
\label{XSFSfig3c_noX}
\end{subfigure}
\caption{Conductance (G) vs. Bias (E) in the $S_1/F_1/N/F_2/S_2$ structure for $D_{F2}=15$ and varying $\phi$. %otv
}
\label{XSFSfig3_JJ_noX}
\end{figure*}

\begin{figure*}[!pth]
\centering
\vspace*{-0.5cm}
\begin{subfigure}[b]{\textwidth}
\includegraphics[angle=-90,width=0.48\textwidth] {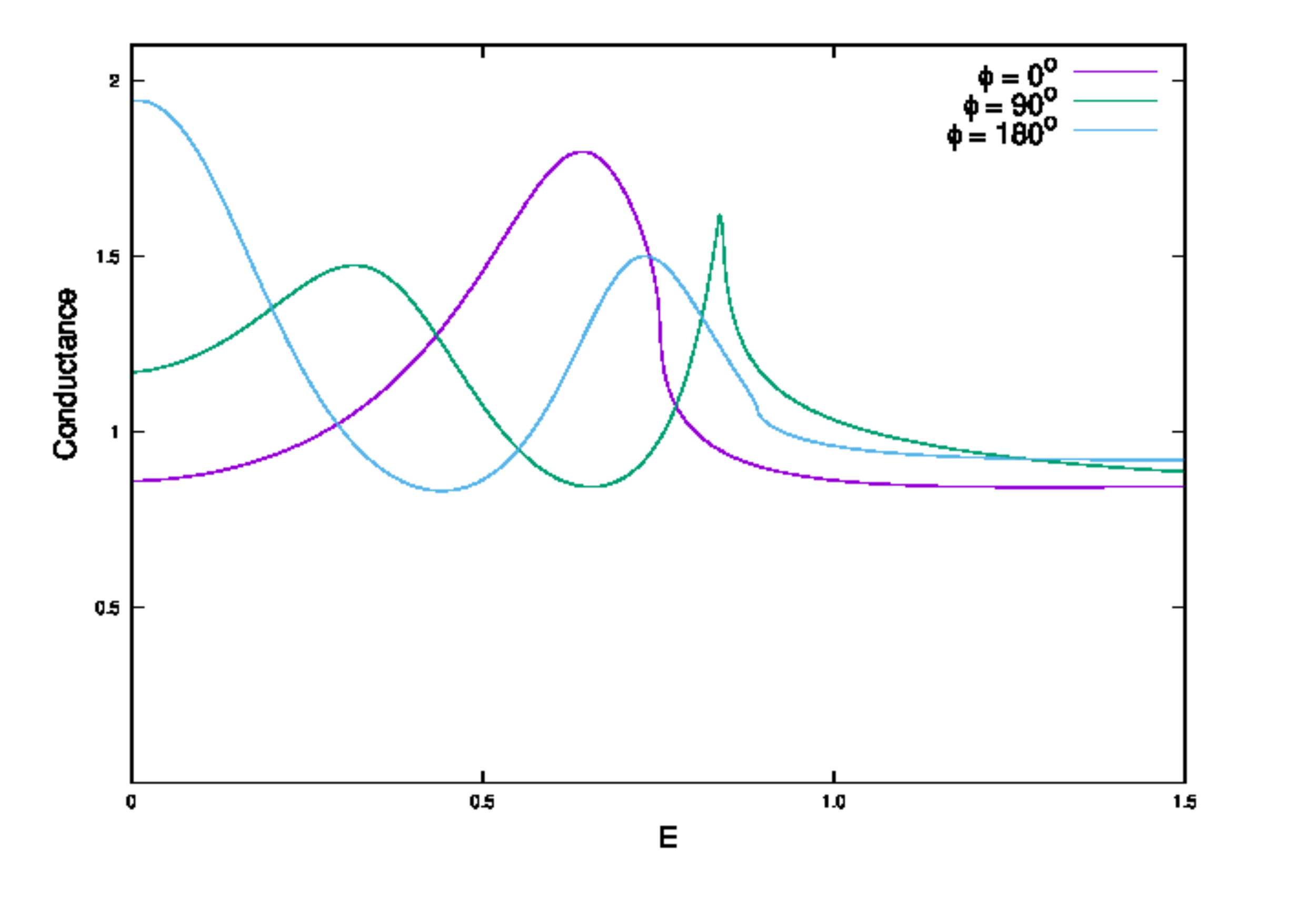}
\vspace*{-0.1cm}
\caption{Single barrier at the $X/S_1$ contact $H_{B,1}=H_{B,2}=H_{B,3}=H_{B,4}=0$, $H_{B,0}=0.5$.
}
\label{XSFSfig3a}
\end{subfigure}
\begin{subfigure}[b]{\textwidth}
\includegraphics[angle=-90,width=0.48\textwidth] {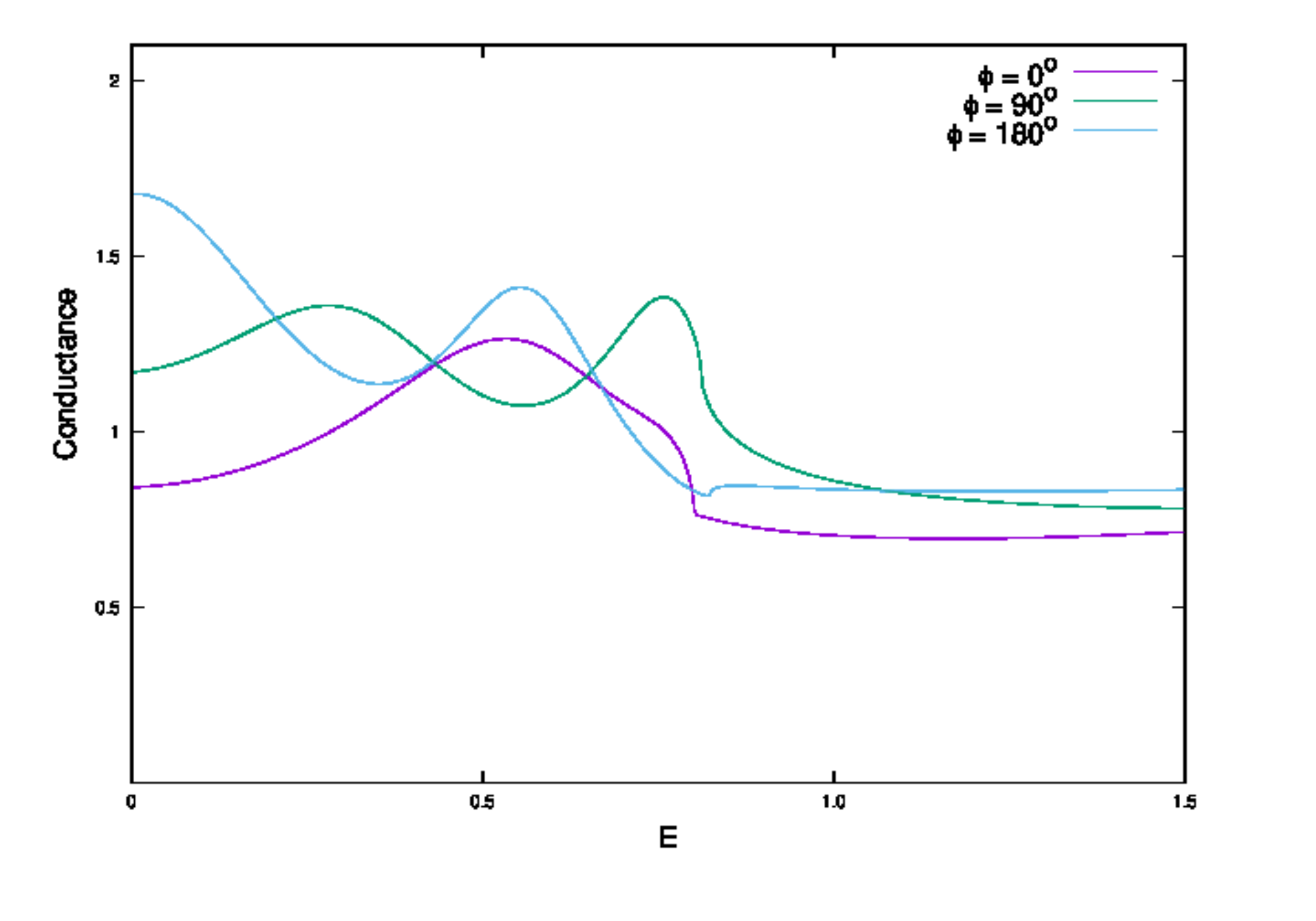}
\vspace*{-0.1cm}
\caption{Barrier at the $X/S_1$ and $F/N$ interfaces $H_{B,1}=H_{B,4}=0$, $H_{B,0}=H_{B,2}=H_{B,3}=0.5$
}
\label{XSFSfig3b}
\end{subfigure}
\begin{subfigure}[b]{\textwidth}
\includegraphics[width=0.48\textwidth] {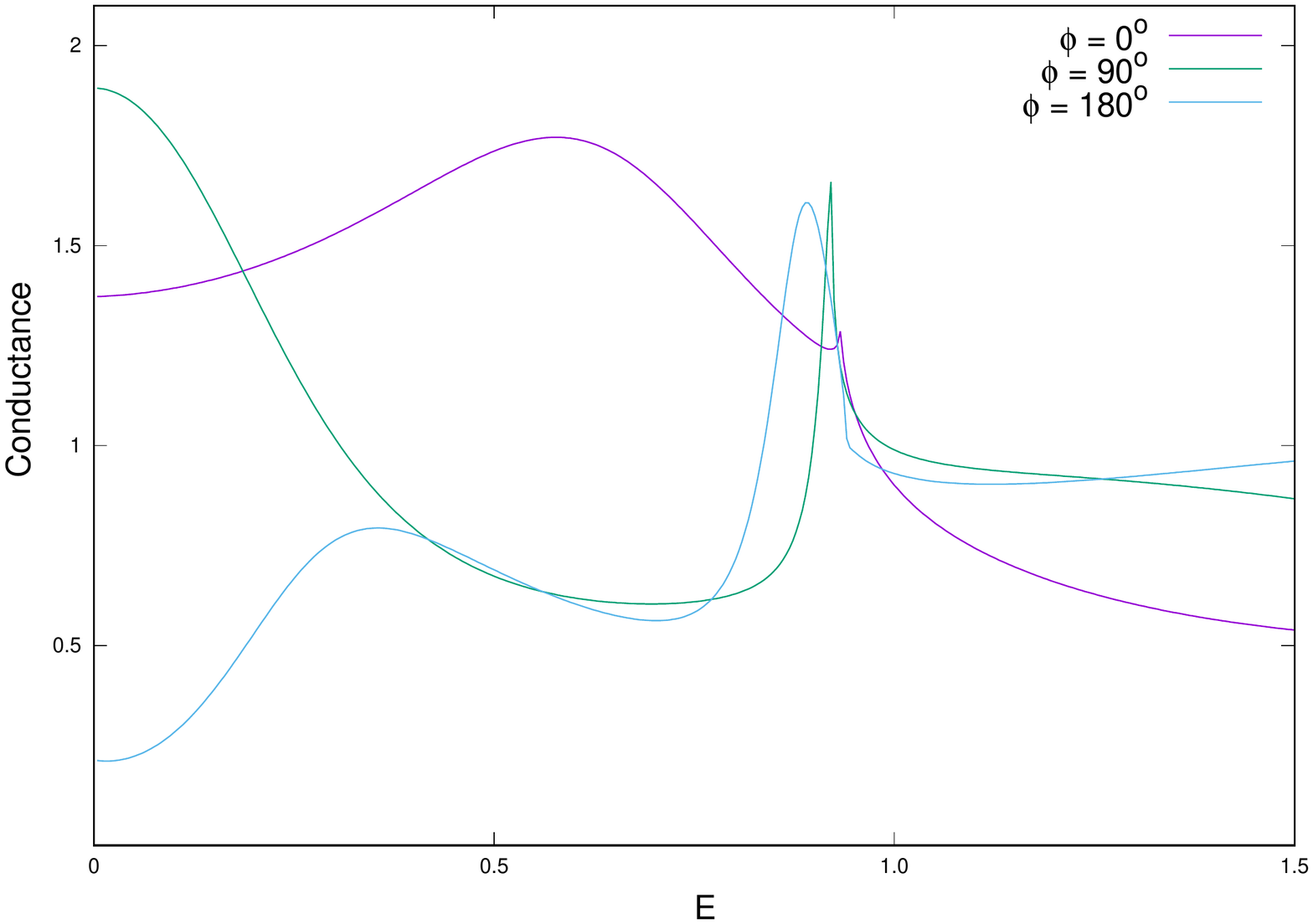}
\vspace*{-0.1cm}
\caption{Barrier at the $X/S_1$ and $F/S$ interfaces $H_{B,2}=H_{B,3}=0$, $H_{B,0}=H_{B,1}=H_{B,4}=0.5$.
}
\label{XSFSfig3c}
\end{subfigure}
\caption{Numerical results for the %otv3
conductance (G) vs. Bias (E) in the $S_1/F_1/N/F_2/S_2$ structure for $D_{F2}=15$ and varying $\phi$. Interfacial barriers as indicated. %otv
% The angular dependence closely resembles the thickness dependence for one-quarter of a period.
}
\label{XSFSfig3_JJ}
\end{figure*}

\begin{figure*}[!pth]
\centering
\vspace*{-0.5cm}
\begin{subfigure}[b]{\textwidth}
\includegraphics[width=0.48\textwidth] {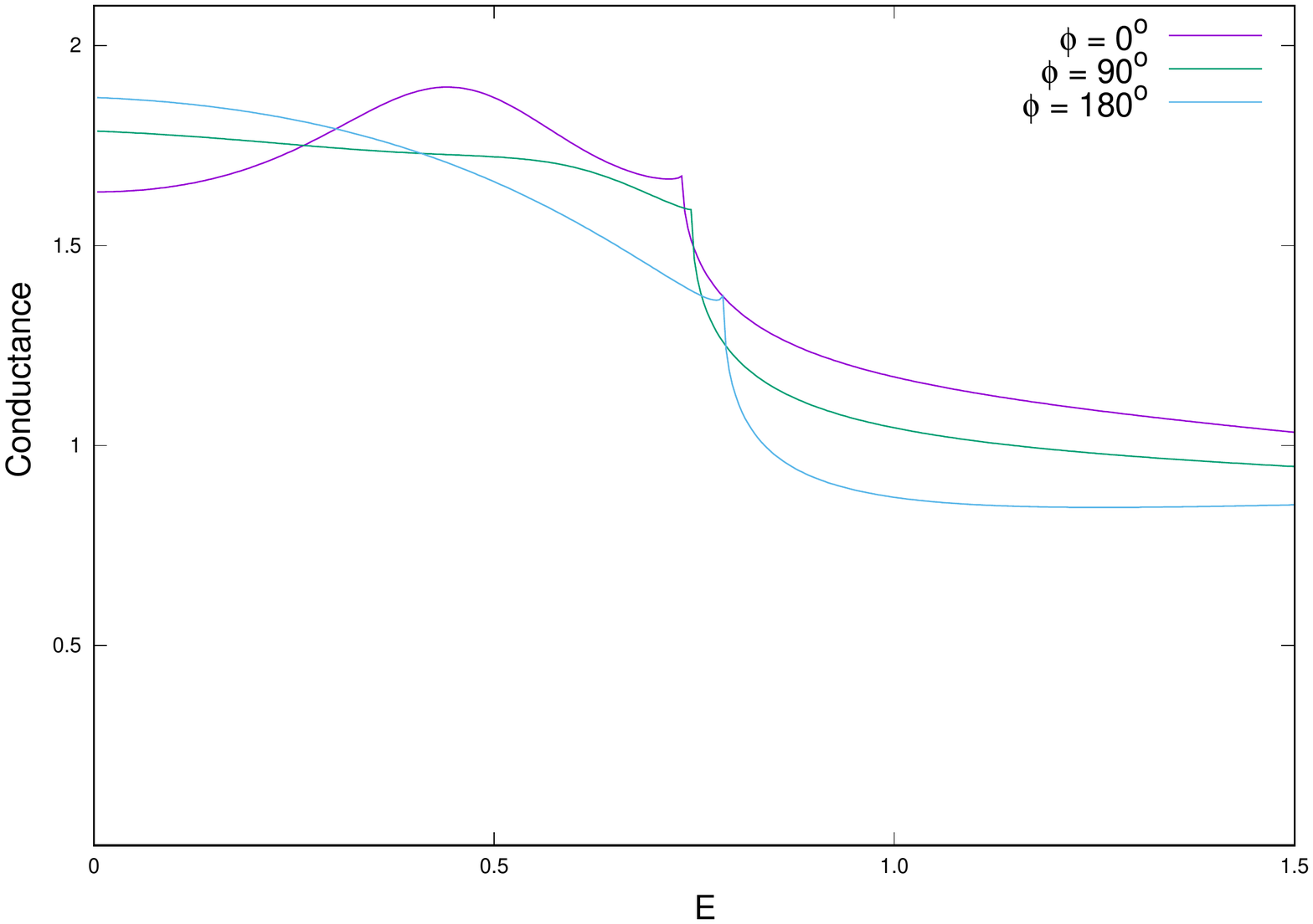}
\vspace*{-0.1cm}
\caption{Barrier at the $F/N$ interfaces $H_{B,0}=H_{B,1}=H_{B,4}=0$, $H_{B,2}=H_{B,3}=0.5$.
}
\label{XSFSfig4b_noX}
\end{subfigure}
\begin{subfigure}[b]{\textwidth}
\includegraphics[width=0.48\textwidth] {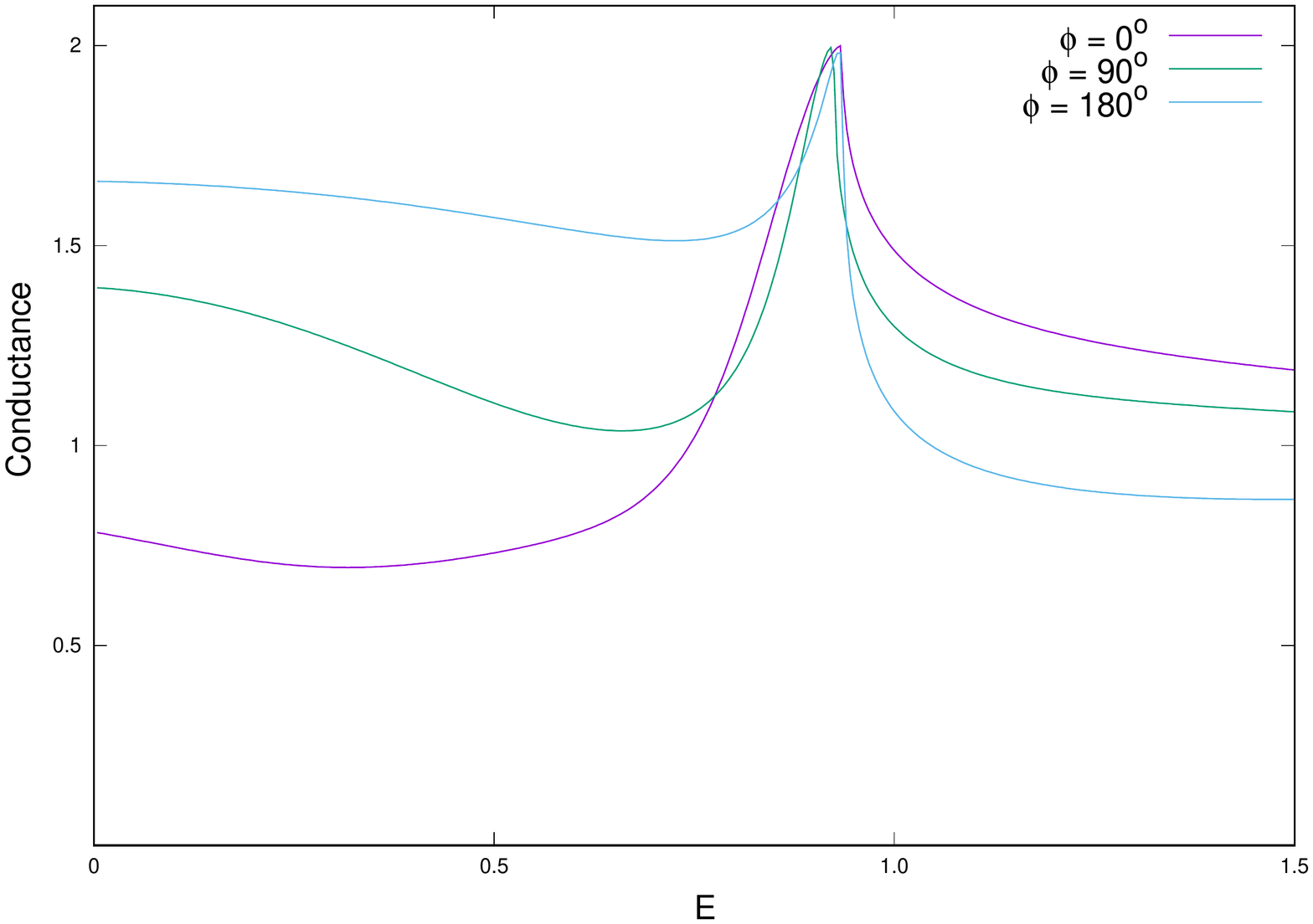}
\vspace*{-0.1cm}
\caption{Barrier at the $F/S$ interfaces $H_{B,0}=H_{B,2}=H_{B,3}=0$, $H_{B,1}=H_{B,4}=0.5$.
}
\label{XSFSfig4c_noX}
\end{subfigure}
\caption{Conductance (G) vs. Bias (E) in the $S_1/F_1/N/F_2/S_2$ structure for $D_{F2}=26$ and varying $\phi$ with a transparent $X/S$ interface.
}
\label{XSFSfig4_JJ_noX}
\end{figure*}

\begin{figure*}[!pth]
\centering
\vspace*{-0.5cm}
\begin{subfigure}[b]{\textwidth}
\includegraphics[angle=-90,width=0.48\textwidth] {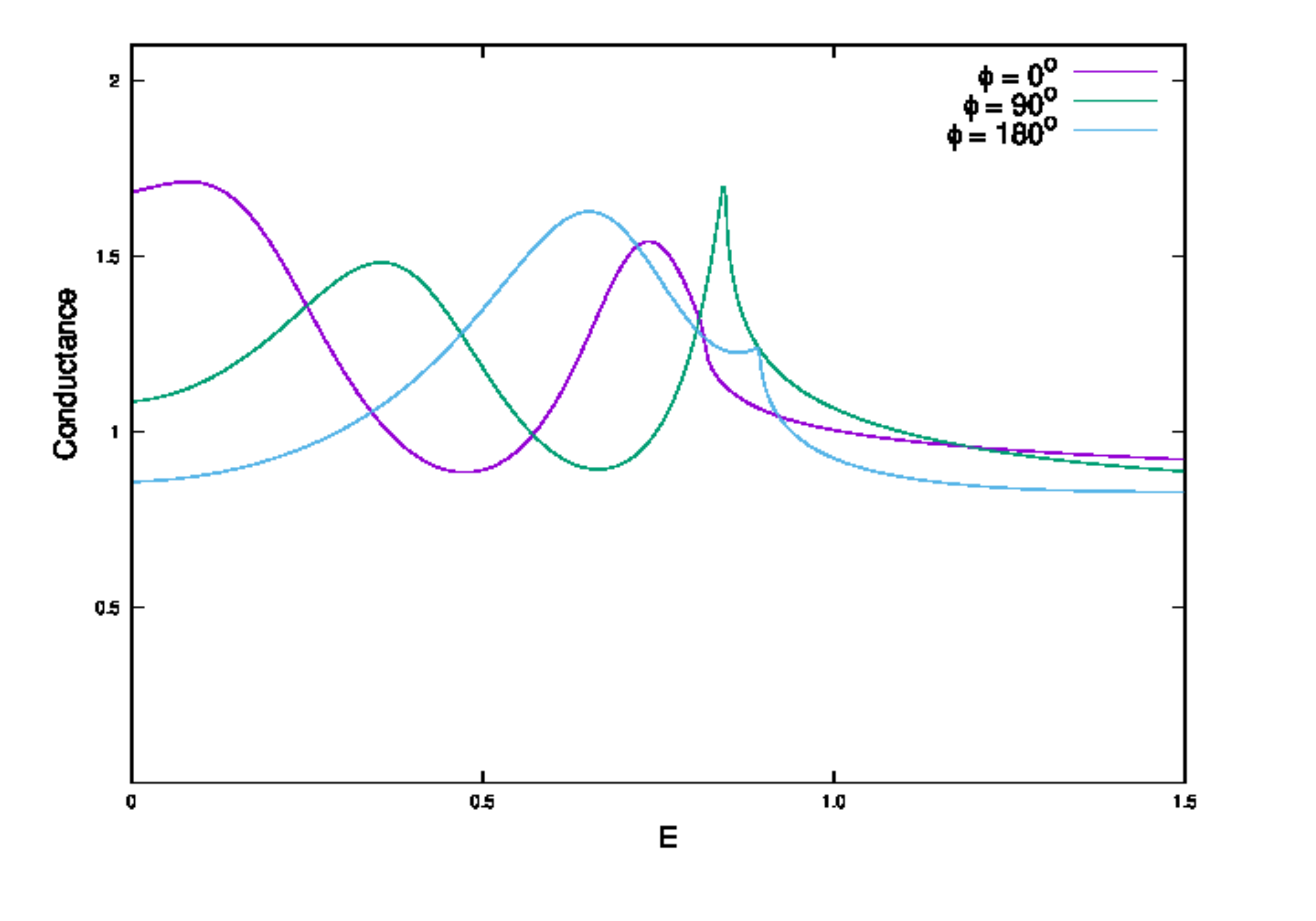}
\vspace*{-0.1cm}
\caption{Single barrier at the $X/S_1$ contact $H_{B,1}=H_{B,2}=H_{B,3}=H_{B,4}=0$, $H_{B,0}=0.5$.
}
\label{XSFSfig4a}
\end{subfigure}
\begin{subfigure}[b]{\textwidth}
\includegraphics[angle=-90,width=0.48\textwidth] {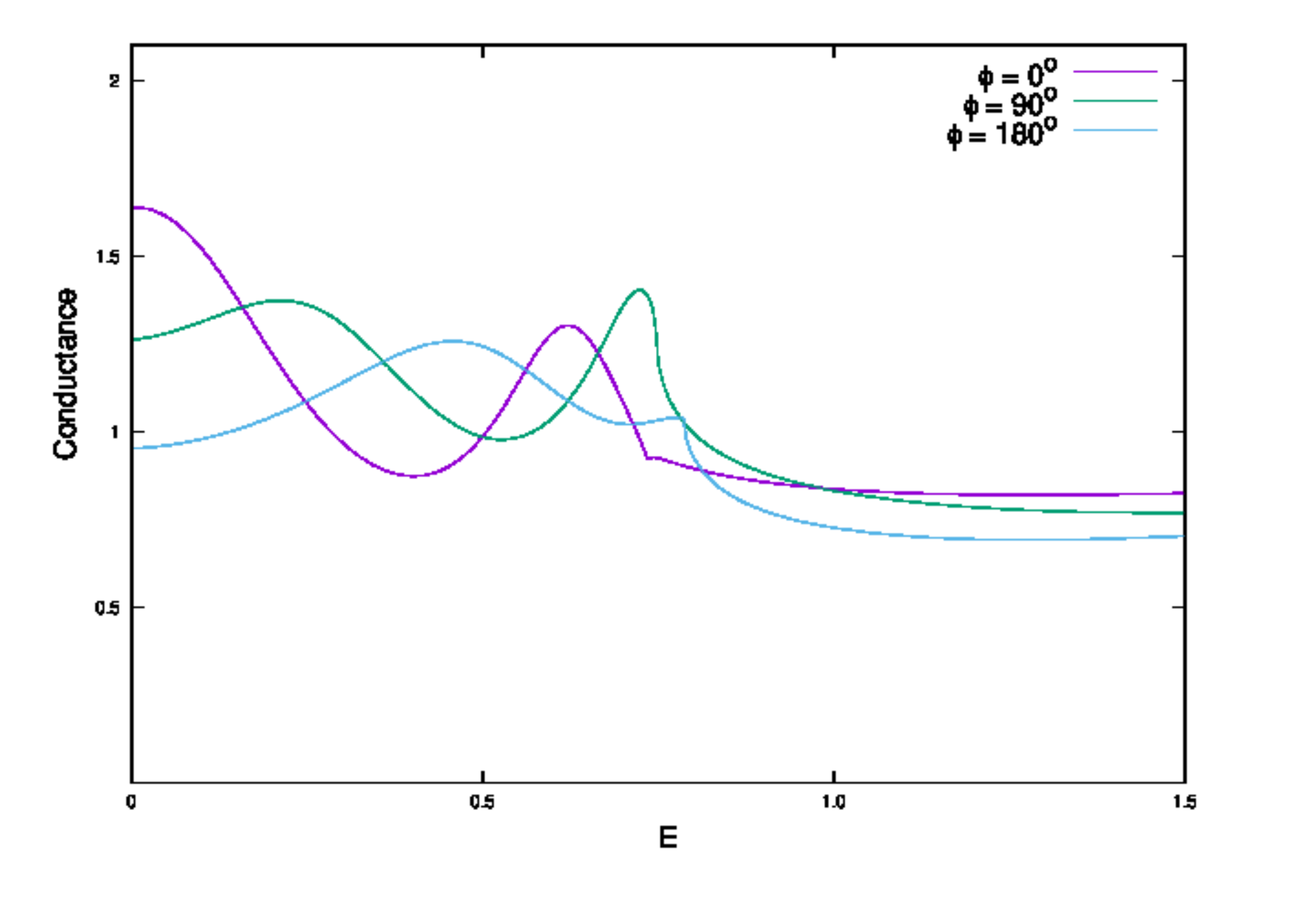}
\vspace*{-0.1cm}
\caption{Barrier at the $X/S_1$ and $F/N$ interfaces $H_{B,1}=H_{B,4}=0$, $H_{B,0}=H_{B,2}=H_{B,3}=0.5$
}
\label{XSFSfig4b}
\end{subfigure}
\begin{subfigure}[b]{\textwidth}
\includegraphics[width=0.48\textwidth] {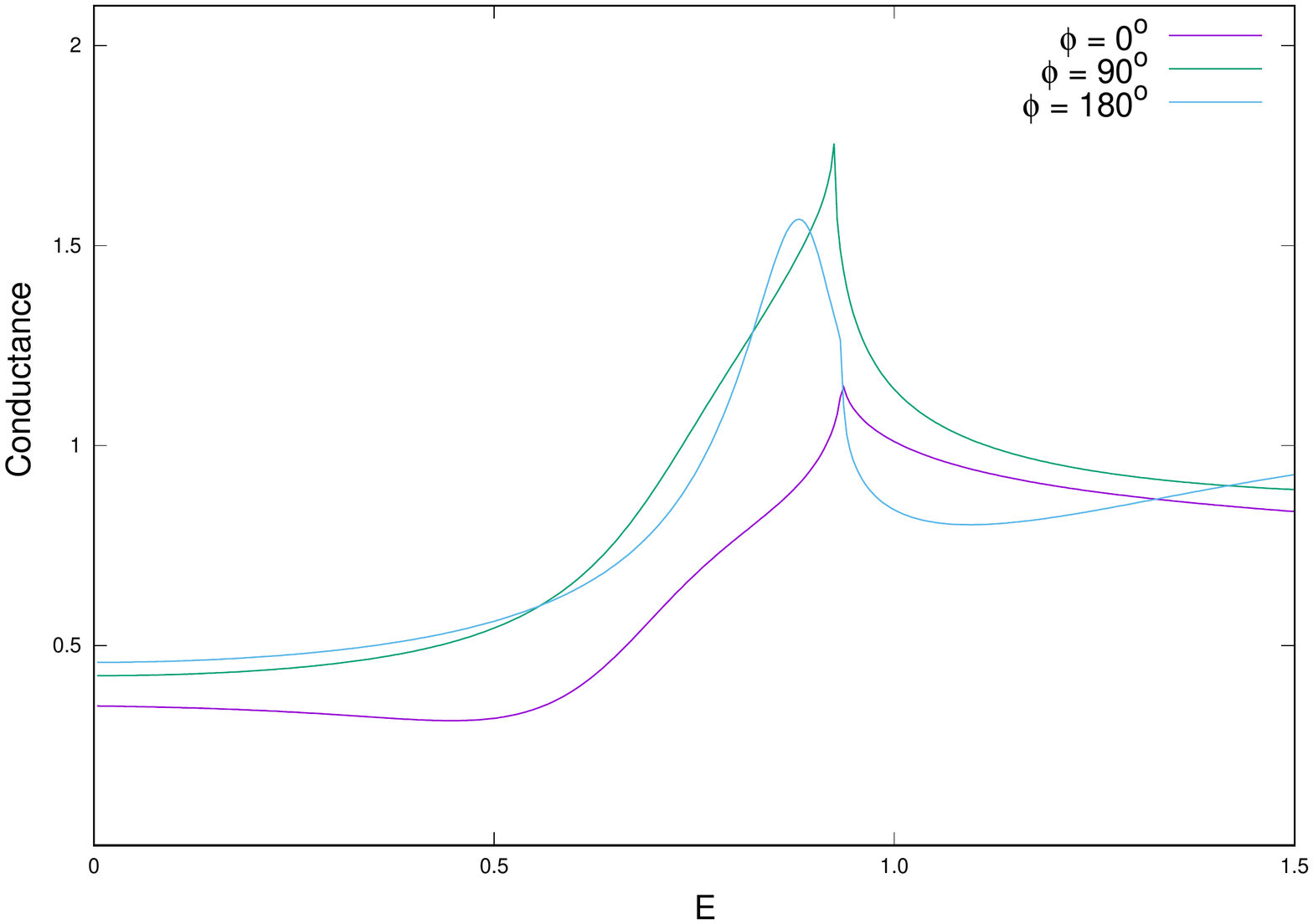}
\vspace*{-0.1cm}
\caption{Barrier at the $X/S_1$ and $F/S$ interfaces $H_{B,2}=H_{B,3}=0$, $H_{B,0}=H_{B,1}=H_{B,4}=0.5$.
}
\label{XSFSfig4c}
\end{subfigure}
\caption{Numerical results for the %otv3
conductance (G) vs. Bias (E) in the $S_1/F_1/N/F_2/S_2$ structure for $D_{F2}=26$ and varying $\phi$. Interfacial scattering at $X/S_1$ contact $H_{B,0}=0.5$. %otv
% The angular dependence closely resembles the thickness dependence for one-quarter of a period.
}
\label{XSFSfig4_JJ}
\end{figure*}

Much of the interest in the spin-valve Josephson structure arises from
its putative capability %otv
to store information in the relative orientation of the magnetization in the $F$ layers. The
angular dependence of the conductance constitutes a valve effect in the system.
In the superconducting spin valve structure ($F/N/F/S$) studied in Ref.~\onlinecite{spinsplitpaper}, 
we found a large valve effect in the subgap conductance for certain thicknesses of the $F_2$ layer.
We aim here %otvr
to determine the angular dependence of the $S_1/F_1/N/F_2/S_2$ structure and 
the viability of the valve effect found. To do this, we analyze $G$ for
two of the thicknesses
plotted in Sec.~\ref{Results_XSFS_df2}, $D_{F2}=15$ and $D_{F2}=26$. 
We choose these two thickness values because they are separated by a value of $\pi/2h\approx11$, one quarter of the full periodicity with $H_{B,0}=0.5$ and % otv3 do not know about all \neq 0
half the full periodicity for $H_{B,0}=0$. 
We will then compare the angular dependence of $\phi$ with the spatial dependence found in Figs.~\ref{XSFSfig2_JJ_noX} and \ref{XSFSfig2_JJ}.
For each thickness, we will study the $H_{B,0}=0$ case (Figs.~\ref{XSFSfig3_JJ_noX} and \ref{XSFSfig4_JJ_noX}) and the $H_{B,0}=0.5$ case (Figs.~\ref{XSFSfig3_JJ} and \ref{XSFSfig4_JJ}).
Then, within each figure we compare the effects that the other interfacial barriers have on the angular dependence of the conductance.

In Figs.~\ref{XSFSfig3_JJ_noX} and \ref{XSFSfig3_JJ}
we plot the conductance for $D_{F2}=15$ and display its $\phi$ dependence. %otv
The $\phi=0$ results are the same as those in Figs.~\ref{XSFSfig2_JJ_noX} and \ref{XSFSfig2_JJ} 
respectively. In Fig.~\ref{XSFSfig3b_noX} we plot the conductance for $H_{B,2}=H_{B,3}=0.5$. 
We see a peak in the conductance at low bias for $\phi=0$ which then transitions to a subgap peak at $\phi=180^\circ$. This is reminiscent of
the thickness dependence, where the single low bias peak at $D_{F2}=15$ transitions into a subgap peak at $D_{F2}=26$ in Fig.~\ref{XSFSfig2b_noX}. 
This parallel extends to the $H_{B,1}=H_{B,4}=0.5$ case in Fig.~\ref{XSFSfig3c_noX}. At $\phi=0$ there is only a small dip in the conductance near the critical bias. At $\phi=180^\circ$
the low bias conductance drops to a minimum value, similar to the $D_{F2}=26$ 
case in Fig.~\ref{XSFSfig2c_noX}. However, in addition to the dip structure, there appears to be a small
inflection near $E=0.4$ which was not seen for $\phi=0$ in the $H_{B,0}=0$ case. This feature is more similar to those found in Fig.~\ref{XSFSfig2c}.

In Fig.~\ref{XSFSfig3a}
we see a single-peaked conductance at $\phi=0$. At $\phi=90^\circ$ the single 
peak splits into a subgap peak and a CB peak. Then, at $\phi=180^\circ$, the conductance has two subgap
peaks, one at low bias and one just below the critical bias. This angular dependence is also qualitatively the same as the thickness dependence going from $D_{F2}=15$ to $D_{F2}=26$ 
as in Fig.~\ref{XSFSfig2_JJ}.
For more realistic interfacial scattering at the $F/N$ interfaces, such as in Fig.~\ref{XSFSfig3b},
we see the same qualitative features in the angular dependence. However, unlike in the superconducting spin valve case, the introduction of these barriers does not enhance
the valve effect. The peaks decrease in value with increased $F/N$ barrier. If we consider the $F/S$ barriers instead, such as in Fig.~\ref{XSFSfig3c}, we see a very different
angular dependence from that of the other two subfigures. It maintains similarities with the $D_{F2}$ dependence, where we see a complex angular dependence with both a peak and dip
structure. However at $\phi=180^\circ$, the inflection point now forms a small peak which is substantially lower than the critical bias conductance, 
while maintaining a minimum near zero bias.

In general, what we can conclude is that the angular dependence between $\phi=0$ and $\phi=180^\circ$ is quantitatively similar to the thickness dependence
going from $D_{F2}=15$ to $D_{F2}=26$ for the parallel configuration. 
This is a striking result: in the superconducting spin valve\cite{spinsplitpaper}
the angular dependence constitutes a uniformly increasing or decreasing conductance peak going from a parallel to
antiparallel configuration with the position of the peak being dependent on $D_{F2}$ only. In the ferromagnetic Josephson structure, 
the angular dependence does not affect the height of the peaks, but instead the position of the peaks just as with the $D_{F2}$ dependence. 
This allows for an extremely large valve effect for almost {\it any} bias value,
as seen in Fig.~\ref{XSFSfig3a} 
where we see a difference in conductance on the order of the quantum of conductance between $0$ and $180^\circ$ at low biases, and $0$ and $90^\circ$ near the CB.
Just as with the thickness dependence, the $F/S$ barriers have the greatest impact on the angular dependence of the system. The $X/S_1$ barrier allows for a more complex
subgap structure with multiple peaks which oscillate and combine, which is reflected in the angular dependence.

In Figs.~\ref{XSFSfig4_JJ_noX} and \ref{XSFSfig4_JJ} we show the angular dependence for $D_{F2}=26$. 
In the $H_{B,0}=0$ case in Fig.~\ref{XSFSfig4_JJ_noX} we see that the angular dependence closely resembles the angular dependence in Fig.~\ref{XSFSfig3_JJ_noX} for supplementary angles: that  is, %otv
the conductance for $\phi=0$ and $D_{F2}=15$ is similar to the 
conductance for $\phi=180$ and $D_{F2}=26$ and vice versa. This corresponds to the $\pi/h$ periodicity, where
the thickness difference between $15$ and $26$ represents half the period in the thickness dependence. Between the parallel ($\phi=0$) and antiparallel ($\phi=180^\circ$)
configuration, the phase advances by $\pi$.%otv explanation?
%EMpaper2 - I'm not sure why the phase advances with phi the same way it does with DF2.
%Going from 0 to 180 is equivalent of DF increasing by pi/h, regardless
%of the H_0 barrier as mentioned below

We see a similar pattern in the angular dependence in the $H_{B,0}=0.5$ cases as well.
The conductance for $\phi=0$ is the same as in Fig.~\ref{XSFSfig2_JJ} where at $D_{F2}=26$ we see
two peaks: one at low bias and one within the subgap region. For $\phi=90^\circ$ we see the two peaks shift to the right, with the 
higher $E$  peak moving into the critical bias. Finally %otv
at $\phi=180^\circ$ the conductance has merged into a single peak just below the critical bias. 
This is the same behavior as seen in Fig.~\ref{NSCfig4_JJ}, except that the periodicity in the thickness is $2\pi/h$. Indeed, the angle $\phi$ advances
the phase of the overall oscillatory spin-split behavior by $\pi/2$ 
when going from a parallel to an antiparallel configuration. 
In the $F/N/F/S$ case, only for certain ranges of thicknesses would the valve effect be noticeable (when the peak was in the middle of the subgap region). 
We see now that in the $S/F/N/F/S$ structure, the peaks change in position 
with $\phi$. This means the valve effect is apparent for any thickness, as any minimum found
at $\phi=0$ will become a maximum when the magnetization is rotated by a certain angle $\phi$. We also note that the $X/S_1$ barrier doubles the effective periodicity
in the conductance subgap features, but the angular dependence between parallel and antiparallel advances the phase of the thickness by the equivalent $\pi/h$ 
wavelength in both cases.

In Sec.~\ref{JJconc} we summarize our results and how they may apply to real devices and experiments.

%%%%%%%%%%%%%%%%%%%%%%
\section{Conclusion}
%otv very ligth edit only
\label{JJconc}
In this paper we have analyzed the 
quasiparticle conductance
$S_1/F_1/N/F_2/S_2$ ferromagnetic Josephson structure using a numerical approach.
Our analysis is in the ballistic limit, and it includes interfacial scattering characterized by delta-function barrier parameters. 
We have included a normal metal contact $X$, with interfacial imperfections, 
which simplifies the calculation of the conductance via the BTK method.
In the calculation of the pair amplitude we use a self consistent method that allows both superconductors
to have an independent phase. We found that the total phase difference in equilibrium is either $0$ or $\pi$. 
%In this study, we have focused only on the quasiparticle conductance. 

To better understand the
numerical results, we have used an analytic approximation for both the 
$N^\prime/N/S$ and $S/N/S$ systems. In this approximation, we assumed a one-dimensional multilayer
with a constant pair potential $\Delta_0$. We also assume an imperfect normal metal contact (with interfacial scattering) for the $S/N/S$ system
as well as interfacial scattering at the $N^\prime/N$ interface for the $N^\prime/N/S$ system. 
We found that for large thicknesses $D_N$ 
the conductance forms  new peaks
 at the critical bias which we call the resonance peaks. In the $N^\prime/N/S$ case the peaks form at equally spaced
intervals $\lambda_n$ which we call the resonance thicknesses, with harmonic number $n$. 
For higher ordered harmonics ($n\ge1$) there exists multiple peaks which are also evenly spaced between
the zero bias and critical bias conductance. We determined in Sec.~\ref{JJanalyticSub} that these resonances
are due to the interference of Andreev reflected particles at the $N/S$ interface with those reflected at the $N^\prime/N$ interface with non-zero
scattering barriers. At higher harmonics ($n\geq1$) the conductance is oscillatory just above the critical bias 
and slowly decays at the same rate for all harmonics. The frequencies of these oscillations are approximately proportional to 
the harmonic number $n$. In the $S/N/S$ case we found two resonance behaviors: ``even'' and ``odd''. The even harmonic resonances are the same 
as those in the $N^\prime/N/S$ case for even values of $n$, but the odd harmonics have an additional term that depends on the ratio of $D_{S1}/\Xi_0$
(see Eq.~(\ref{ResonanceJJ})). This term reduces the resonant thickness of the odd conductance peaks. The oscillatory conductance above the gap
is also shifted by this $S_1$ thickness dependence in the odd harmonic thicknesses.

We then applied our analytic approximation to the 
ferromagnetic $N/F/S$ and $S/F/S$ systems. 
In Ref.~\onlinecite{spinsplitpaper} we studied the
spin-split conductance of the $N/F/S$ system, where the conductance peak oscillates between the critical bias and near zero bias for varying 
thicknesses of the $F$ layer over a wavelength of $\pi/h$. 
We did so for only small thicknesses of $D_F$, just above the $n=0$ harmonic. In this paper,
we studied the same effects on the $n=1$ harmonic, where there are two conductance peaks. Both peaks
oscillate in position together between the subgap region and the zero bias conductance for the low bias peak, and the critical bias and subgap region for the higher bias peak. 
Between those two thickness
values, each peak splits, resulting in multiple subgap peaks in the conductance. This also applies to the $S/F/S$ case. 
In our analysis of the Josephson structure, we saw that for even relatively small values of the $F$ layer thickness (less than
the coherence length of the superconductor), the conductance displays multiple subgap peaks. This is because the spin-split oscillations
can pull the higher order harmonic peaks into the subgap region since the first harmonic
thickness ($n=1$) is reduced by the presence of the $S_1$ layer. %EMpaper2 - thickness

Armed with this qualitative understanding of the $F$ layer thickness dependence of our $S/F/S$ analytic calculation, we
were then ready to consider
the results for the fully self-consistent $S_1/F_1/N/F_2/S_2$ ferromagnetic Josephson structure. We studied the $F_2$ thickness
dependence in the parallel configuration of the $F$ layer magnetizations ($\phi=0$) for the case of clean $F/N$ interfaces and imperfect $F/N$
interfaces. In some cases we assume a nonzero scattering barrier due to a normal metal
%EMpaper2 - some cases, not all now
 contact $X$, which we found to enhance the conductance peaks
by decreasing the average subgap conductance. 
In our numerical calculations we found the same qualitative features of the subgap conductance
as found in the analytic $S/F/S$ system. 
%EMpaper2 - new below !!!
We carefully considered the barrier dependence of the conductance. We find that the inclusion of an $X/S_1$ scattering barrier $H_{B,0}$ allows for multiple subgap peaks.
For example, by closely observing spin-split oscillation with $D_{F2}$ in Fig.~\ref{XSFSfig2a} we could see how a single 
subgap peak at $D_{F2}=15$ becomes two subgap peaks at $D_{F2}=26$, 
with one peak being near the critical bias and one being at low bias. 
This is in contrast to the $H_{B,0}\neq 0$ case or the
 $F_1/N/F_2/S$ superconducting spin valve where, 
 for similar thicknesses of the $F_1/N/F_2$ layers,
we only saw a single subgap conductance peak.
We found that the inclusion of $F/N$ interfacial scattering does not %otv2 enhance  or
greatly affect the conductance peak structure, but the $F/S$ barriers have
a major impact on the subgap structure by forming dips
in the subgap conductance. These dips do not have the same periodicity as the
peak structure, which can lead to complex features in the subgap such as points
of inflection.

We concluded this work with a study on the the angular dependence of the ferromagnetic Josephson structure. We calculated the conductance
for multiple angles $\phi$ of the relative orientation of the ferromagnetic layer magnetizations (see Fig.~\ref{JJfig0}) 
in the $S_1/F_1/N/F_2/S$ configuration. 
The angular dependence is similar to that on $D_{F2}$. By rotating
$\phi$ between the parallel and antiparallel configuration,  the phase of 
the spin-split conductance oscillations advances by $\pi/2$ 
%EMpaper2 - more edits below!
in the $H_{B,0}\neq0$ case and by $\pi$ in the $H_{B,0}=0$ case.
This is, in both  cases, %otv2
 equivalent to increasing the thickness by $\pi/h$. %otv2 in each case.
This is in stark contrast to the $F_1/N/F_2/S$ structure, 
where the angular dependence was found only in the subgap peak height and {\it not}
in the position of the peaks within the subgap. This allows for a very large valve effect, on the order of the quantum of conductance per
channel, which may prove useful in future spintronic devices.

Although we have learned about many new exciting features unique to the ferromagnetic Josephson structures, there are still many unanswered
questions. For instance, we were unable to analytically determine the odd resonance thicknesses and had to settle for a phenomenological 
approximation. In addition, we have not determined 
how the $S_1/F_1/N/F_2/S_2$ angular dependence is related
to the spin-split conductance oscillations. Many more questions that
could be asked, such as the $S$ and $N$ layer thickness dependencies and even the study
%EMpaper2 - we now have S/F
 of the $\Delta_1\neq\Delta_2$ Josephson structure. 
We also assumed one imperfect contact, and have not studied the effect of two imperfect contacts.
We believe 
that this paper, however, leaves a good foundation and highlights some of the more unique aspects worthy of future study. 
We hope that this work will be useful for future experiments into
ferromagnetic Josephson structures and their application in spintronic devices.

\acknowledgments  The authors thank I.N. Krivorotov (University
of California, Irvine) for many
illuminating discussions on the
experimental issues. This work was supported in part by DOE grant No. DE-SC0014467

\end{document}